# Large-Scale Qualitative Research with AI: Infrastructure, Management and Operation of the Socioscope Data Pipeline

Saadi Lahlou[1,2], Juan Pablo Caicedo[1], Shriya Sekhsaria[1], Valentine Fournand[1], Paulius Yamin[1], Helga Nowotny[3]

[1] Paris Institute for Advanced Study, Paris, France. [2] London School of Economics and Political Science, London, UK. [3] Complexity Science Hub Vienna, Austria.

Corresponding author: Saadi Lahlou - saadi.lahlou@paris-iea.fr

## Abstract

The Socioscope project is a pioneering effort in Large-Scale Qualitative Research (LSQR) collecting comparable, open-ended, multimedia field data on hundreds of cases and using AI to make the material analysable at scale. The domain studied is the food system. The entities documented are the organisations that act in it: farms, processors, distributors, retailers, restaurants; and, at meso level, the actors that shape their environment, such as municipalities, government programmes, banks, NGOs and universities. This paper provides the technical reference for how the resulting data Corpus was built and managed to enable AI-augmented analysis. It describes the data pipeline end to end: the systemic sampling frame; the transaction grid used to capture each initiative's relations within the food system; the social contract that rewards participating interviewees, aiming to sustain access; the operational chain from scouting to interviews, including their uploading, transcription, translation, quality control and curation; the provenance rules (originals are immutable, every transformation is logged); and the installation of equipment, personnel and processes, including ethics and GDPR compliance. In its first phase (2023–2026) the pipeline produced 686 documented cases from 31 countries: some 1,430 hours of recordings, about 450,000 speech turns, and 12.6 million words of transcript. We report costs, metrics, lessons learned and limitations, so that other teams can reuse, adapt, and improve the Socioscope methodology.

Keywords: Large-Scale Qualitative Research; data pipeline; qualitative methods; artificial intelligence; data provenance; research infrastructure; sustainability transitions; food systems.

# 1 Big data, deep data, AI, Large-Scale Qualitative Research and the Socioscope project

This paper describes the Socioscope project, an experiment in Large-Scale Qualitative Research (LSQR). We share our experience for those who would take the same path.

Large-Scale Qualitative Research is a new method for social science research. Typically, quantitative research preformats the data collected in order to process a great number of statistical objects, while qualitative research observes in great, open-ended detail a few cases at a given time. The Socioscope is designed to combine the advantages of both approaches: scale and depth. It investigates many cases through open-ended data collection with the same protocol, leveraging artificial intelligence to extract information in a comparable manner. As AI becomes widely available, we expect Large-Scale Qualitative Research (LSQR) to become a common methodology.

The Socioscope is a LSQR research instrument and methodology built to explore systemic societal change, leveraging AI. Much as astronomy changed with the use of telescopes, the Socioscope is an attempt at the equivalent for social science research: an instrument adapted to better analyse its objects of interest, here the economic entities and their relationships in complex social systems. The system we analyse as a testbed (the food domain) is broad: it comprises many entities (agents, organisations, institutions, devices). It is also deep: these entities are themselves complex subsystems, linked by many kinds of interaction. To address this material, the Socioscope collects deep, comparable data on hundreds of entities and formats these data into “cases” to be studied together.

As we finally start to analyse and get results from this mass of empirical material we are now convinced qualitative research at scale is worth the effort. A large body of comparable open-ended cases allows things a dozen cases do not. Classifications, typologies and ontologies can be built, and these become models, which can later feed simulations. A hypothesis formed on a few cases can be tested on hundreds. The coding grids themselves can be tested for relevance and robustness, and a grid built on part of the corpus can be tested on the rest. Cultural particularities become visible, individual bias in collection and interpretation can be controlled. Different types of actors can be compared with the same grid; the micro, meso and macro levels can be related (Nowotny et al., 2026). The distance between what is done, what is said about it and what is wished can be examined across hundreds of actors. The data stay open-ended, so variables and dimensions not envisaged at the start can be discovered, and can then be tested. The corpus also answers questions that were not asked when it was collected, which a preformatted instrument cannot do. This is what the analysis in progress shows; it will be reported in separate papers. We have not yet explored the full potential of LSQR.

AI is also useful on the research itself, not only on the data. A project of this size accumulates decisions, definitions and exceptions over years and dozens of people. An AI assistant that keeps the record, session minutes, a log of analytical decisions, change logs on each instrument, the state of each case, keeps the group coherent and lets a newcomer query the project instead of reading it. It can help recover documents in the mass of the Corpus. This is provenance applied to the work and not only to the material. Coordination, not storage or compute, is also what has historically kept qualitative research small.

But all this comes at a cost. LSQR is not just scaling up qualitative research; nor is it quantitative surveys with longer and open-ended questionnaires. It is a break with the small-scale, cottage-industry approach, and also with classic quantitative research; several of the differences we experienced were unexpected. It comes with a philosophy of data collection based on participation of the data sources, and a different type of interpretation process, with recursive abduction. It removes some constraints and allows bolder research strategies; it also brings new constraints and higher costs in data collection and management. It requires a formal research infrastructure, which is rather new in this type of social science. These requirements are better explained with a concrete example.

This paper presents the Socioscope's data collection process, and takes stock of LSQR in actual use at scale. We had exceptional freedom to experiment, thanks to generous basic research funding by the NOMIS Foundation. It enabled us to make mistakes, explore dead ends, and build operational solutions.

To research systemic change at societal level, the Socioscope has opted to take food as its test domain: the production, transformation, transport, storage, retail, preparation, consumption, waste management of food, and the infrastructure and organisations that support them. The main research questions are "how do transitions towards greater sustainability happen?" and "how does the food system evolve?". To address them, we wanted to observe what those actors who push the system towards becoming more sustainable do 'on the ground'. Therefore, the Socioscope documents "initiatives" that try to make the food system more sustainable, like new forms of organic farming, low-footprint transformation and distribution, or circular-economy models. It also follows the action of entities at the meso-level; which are stakeholders such as municipalities, financial institutions, universities, governmental programmes and other entities who actively intervene for change.

In its first phase, data collection (2024-2026), and following a proof of concept, the Socioscope has to date documented over 700 initiatives across 37 countries on five continents (686 of them released in the Corpus used here for statistics, version 0.1.0, covering 31 countries, as others are still in final processing), with the ambition of producing findings about systemic transition and, just as important, a reusable methodology for studying transition processes (Lahlou et al., 2024). The first data collection phase is now complete, analysis is work in progress. Pages presenting each participating initiative - a short abstract of its case and the video made for it - are published on the project public website, www.thesocioscope.org.

The main focus of this description of the Socioscope's methodology as LSQR is the data pipeline: how data sources are identified and managed, how the technical chain is designed and run, how the workforce that feeds and curates it is organised, and how the data move from collection through curation to analysis. The paper is organised as follows: the remainder of this section situates Large-Scale Qualitative Research with respect to big data and deep data. The Socioscope Data Pipeline (Section 2) presents the design principles: the choice of data sources, the transaction grid used for data extraction, the Corpus and the Gate that guard it, and the social contract with participating interviewees. Data enrichment and release for analysis (Section 3) describes how the Corpus is designed, how provenance is preserved, and how data are enriched and released for analysis. The Socioscope installation (Section 4) details the infrastructure required: equipment, personnel and processes, including ethical oversight. Data journey step by step (Section 5) walks through the operational sequence of a case, from scouting to the fulfilment of the social contract. Variations and costs of the pipeline (Section 6) compares three configurations (human-only, tech-maximal, hybrid) and their economics. The Socioscope pipeline implementation (Section 7) reports what the pipeline produced in two and a half years. Lessons learned and limitations (Section 8) takes stock, and the Conclusion (Section 9) draws a more general outlook beyond our project. The appendices provide a checklist of design recommendations for other research teams attempting LSQR (Appendix 1), the step-by-step pipeline with the human/AI/hybrid comparison (Appendix 2), the structure of a case folder (Appendix 3), pipeline metrics (Appendix 4), and the project workforce in Phase 1 of the project (Appendices 5 and 6), and the contents of the data-collection protocol (Appendix 7). The full data-collection protocol (55 pages) is available from the authors on request.

### 1.1 **Large-Scale Qualitative Research: Combining breadth and depth**

Research on societal and sustainability transitions has grown into a substantial field over the past two decades (Markard, Raven, & Truffer, 2012). Like all systemic research, it remains caught between two pitfalls. At one extreme is the "model approach" with generic typologies and models, many abstract concepts theorised at considerable distance from the ground and resting on aggregate data. At the other end is the "case study": rich, in-depth investigations whose findings are difficult to generalise, because each case is specific, cases are few, or too unevenly distributed to compare. Behind this lies the dilemma of "big data" versus "deep data" which limits the study of complex societal systems because of their size and dynamics.

The Socioscope's methodology combines what was long held to be incompatible: the breadth of large-scale comparison and the depth of qualitative case studies.

Big data offers large samples of predominantly economic and other administrative statistics, often gathered for other purposes, or collected with questionnaires, automatic measurement devices (e.g. logs, sensors), or traces of operations by users (e.g. buying acts, social media posts). Deep data, or "thick data" as Geertz called them (Geertz, 1973), by contrast, are collected with a specifically tailored intervention e.g. on-site interviews or ethnographic observations resulting in detailed case studies which resist comparison.

Each tradition has codified its answer in classic textbooks. The quantitative canon urges researchers to maximise the number of observations to gain inferential leverage: cases matter as long as they are numerous, comparable, and sampled to support generalisation (King, Keohane, & Verba, 1994). The qualitative classics defend the opposite position: grounded theory builds concepts from intensive engagement with a small number of sites (Glaser & Strauss, 1967); ethnography seeks the "thick description" whose value lies in contextual density (Geertz, 1973); and the case-study tradition argues that the power of the case is exemplary and context-dependent rather than statistical (Flyvbjerg, 2006; Yin, 2018). Small (2009) diagnoses the resulting impasse: asking "how many cases do I need?" for a field study imports a sampling logic foreign to case-based research. Big data has its own warnings: numbers do not speak for themselves, and bigger data are not always better data, because they are shaped by the instruments and platforms that produce them (Boyd & Crawford, 2012).

Social scientists have worked within the big-versus-deep constraint for a long time, and with hard work, ingenuity and trade-offs have produced important findings, theories and methods. The opposition between quantitative and qualitative camps is past. But the dilemma, and the ambition to overcome it, are not.

There has been no lack of attempts. Qualitative Comparative Analysis formalises case study knowledge into configurations of conditions, so that intermediate numbers of cases (too many for narratives, too few for statistics) can be compared systematically (Ragin, 1987). Mixed methods designs combine the two logics by sequencing or nesting them, for example by running intensive case studies inside a statistical frame (Lieberman, 2005; Tashakkori & Teddlie, 2010), or starting with qualitative research to design questionnaires. These strategies either compress the qualitative material into pre-defined conditions at the point of entry, or juxtapose the two logics without merging them in a single instrument. More recent developments are closer to the Socioscope methodology: they treat qualitative material as data at scale.

A tradition of qualitative data archiving and reuse, institutionalised by Qualidata and its successors in the UK Data Service and codified in good-practice guides (Corti, Van den Eynden, Bishop, & Woollard, 2019), has built the machinery to preserve qualitative material and make it findable and reusable. Computational social science has shown how much can be learned from big data but also exposed their limits and the institutional obstacles to using them well (Lazer et al., 2009; Lazer et al., 2020). On the analysis side, the “big qual” current has developed methods for working across large assemblages of archived qualitative studies: the breadth-and-depth method, for instance, alternates computational surface mapping of a large corpus with deep interpretive “test pits” sunk at selected points (Davidson, Edwards, Jamieson, & Weller, 2019). Recently, data mining has used statistical techniques (Manning & Schütze, 1999; Grimmer, Roberts, & Stewart, 2022), or AI (for a review, see Ziems et al., 2024) to process large amounts of text, usually cropped from digital sources. The Socioscope shares the premise of this work: qualitative material can be treated as data at scale and remain qualitative. But it inverts the workflow. Big-qual analysis assembles heterogeneous legacy studies and compensates for their differences afterwards. The Socioscope builds comparability upstream: it creates the data it uses instead of copying them from an existing source. Every case is collected with the same instrument, under the same protocol, and put through the same pipeline. We ventured into doing this because we knew that we could indeed process that mass of material using AI; which our predecessors could not hope.

### 1.2 AI unlocks qualitative data processing at scale

Before AI, qualitative research could not easily be scaled up because analysing large amounts of qualitative data required extensive human time, manual coding, and close interpretation of meaning (Miles, Huberman, & Saldaña, 2020; Richards, 2015). Larger datasets increased the difficulty of maintaining consistent coding, managing complexity, and preserving the contextual depth needed for qualitative analysis (Bazeley, 2013; Braun & Clarke, 2021). Existing software, e.g. CAQDAS (Computer-Assisted Qualitative Data Analysis Software) could help organise data but not replace the role of researchers in identifying patterns and interpreting meaning, which limited the size of studies that could be practically conducted (Johnston, 2006; Friese, 2016). As a result, qualitative research was often limited to smaller samples to balance analytical depth with available time and resources (Brower et al., 2019).

Generative AI changed this situation. The Socioscope came at the right time. It uses large language models (LLM) to mine a large data set, as text mining does with pre-existing data. But it goes further: AI also assists the collection of original qualitative data, designed from the outset for systemic analysis at large scale. This is close to computational grounded theory, which alternates algorithmic pattern detection with close human reading (Nelson, 2020). Our project goes one step further, from existing text to a corpus purpose-built for comparison. Each case is described not only in classic terms (what initiatives are, what they do) but in its relations with the rest of the system. This is done through open interviews that leave freedom of expression to the interviewee, while covering a broader range of significant activity (see below the “transaction grid”).

AI brings several benefits to LSQR: it can process large volumes of data, transcribe interviews with advanced speech recognition (Chowdhary, 2020), and identify themes and patterns in text (Verma et al., 2021). Systematic assessments now document LLM annotation performance rivalling trained human coders, including in multilingual settings (Gilardi et al., 2023; Rathje et al., 2024; Törnberg, 2025), and best-practice standards are emerging (Törnberg, 2024; Bail, 2024; Karjus, 2025). It also introduces specific problems: data privacy, transparency and reproducibility of output, perpetuation of biases that are present in the training data, and researcher accountability (Christou, 2023). AI models evolve continuously and are stochastic rather than deterministic. The models used must therefore be documented exactly, even though exact reproduction is not possible.

Despite these limitations, AI supports qualitative research: it saves effort and widens access. As Anis and French (2023) argue, AI should complement human expertise, not replace it. If these limitations are managed, as in the Socioscope’s LSQR methodology, researchers can keep the depth, reliability and ethical integrity of their work.

### 1.3 Collecting data with a systemic approach

The Socioscope rests on the conviction that the quality of the data is essential to good research, that value is in the data (Lahlou, 2025). If properly collected and documented in a data Corpus, the data lend themselves to be later used to answer different research questions. This matters because the research question usually evolves as more material is collected and the problem is better understood. As the research proceeds and questions transform, researchers often discover that the data needed were not collected in the first place. It is therefore advisable to collect the data with a more generic understanding of the system in view, leaving room for research questions other than those posed initially.

This approach also addresses the recurrent problem of data re-use. Re-use is difficult today, for other teams and even for the same one, because collection was tailored to the initial purpose and documentation is vague or partly tacit. Two things are needed: good documentation, which clarifies the potential and the limits of the data and keeps access to the sources open, so that the corpus can later be completed with new data from the same sources; and system-oriented collection, which situates each case in the larger system and records its links with other entities. Changes in these relationships are what one must follow to understand transition processes.

The Socioscope collects multilingual qualitative field data, as transcripts of speech and as visual material on the setting, and formats them into a homogeneous, analysable, extendable corpus. The methodology is organised around a data pipeline that supports tracing and documentation and remains open to enrichment. This requires a research infrastructure covering the three layers of an “installation” (Lahlou, 2017) to produce the data collection: material affordances of physical equipment, embodied competences in the team, and regulation in the form of contracts, rules and processes.

The Socioscope methodology therefore has three components:

1. **Data collection and processing pipeline:** the journey of the data from the scouting and selection of cases to their incorporation in the Corpus, in the right format, with labels and metadata. It includes the management of access, the extraction of data for analysis, and the maintenance of metadata.

2. **Analysis and enrichment:** the extraction of information and analyses from the Corpus. Some outputs are fed back into the Corpus, together with the documentation of the analysis.

3. **Infrastructure:** the installation that supports collection and analysis: physical equipment (computers, recorders, storage, means of communication…), competent personnel (for data collection, analysis, including support personnel, HR…), and procedures (contracts, protocols, role definitions, quality control, management…).

As in any research, purpose and research questions are defined first, and they determine the kind of data and the data sources. The collection process is then designed. The research strategy may change later, but it must always be stated explicitly.

This paper describes how the Socioscope methodology was built and how it operates. After a brief description of the research strategy, we present the data pipeline in full, then analysis and enrichment in less detail, then the infrastructure the processes require. The appendices give more detail.

## 2 The Socioscope Data Pipeline

Every research project must choose its data sources and the process of extraction. The data must contain the information needed to answer the research question.

A first question (where?) is how to define the unit of analysis and which data sources to sample (what type of entities are good sources of information for our problem, and which ones in particular should we select). A second question (what?) is which data we need to collect on these units to address our research question. A third question (how?) concerns the conditions under which the desired data can actually be collected from the sample of sources. The sections below say how we handled each.

## 2.1 Choosing the data sources

### 2.1.1 Nature of data sources

The sampling unit, which we call an “entity”, is the individual source of data. The Socioscope is interested in organisations relevant to how the food domain works and evolves: a farm, a fast-food chain, a frozen-food company, and so on.

A systemic approach obliges us to identify and situate precisely the entity observed, as the unit of analysis, in order to understand the transition processes at work. The Socioscope studies the evolution of the food system in a given direction, namely moving towards greater sustainability. We want to understand what processes, devices, mechanisms and factors operate when actors try to move the system towards greater sustainability. We therefore focus on agents who explicitly try to move the system in that direction. We call them “initiatives”. Initiatives are usually collective undertakings, even when individuals play an important role, and take various organisational forms.

We first identified initiatives at the micro-level, to observe what happens on the ground, by scouting and then screening them against our selection criteria (see Sample selection). Following their interactions with other initiatives, partners and stakeholders led us to include initiatives at the meso-level: municipalities, governmental programmes, banks, NGOs, large corporations, which also become data sources. We meet initiatives in their own habitat, interview them, film their facilities and ask for supplementary material describing their activities, products and relations with the larger system. The collected data are digitized and archived. The data about a data source that has been collected, digitized and archived become “a case”.

For example, AR-068 in Argentina, is such an initiative. AR-068 is part of the Participatory Guarantee System (PGS): an agroecological certification initiative that fosters collaboration between producers, consumers and technical experts. Established in the Andean region, it aims to build trust through a locally developed seal that certifies agroecological practices. Through meetings and guided tours, the PGS shares knowledge, strengthens community ties, encourages producers to convert to agroecology and informs consumers. FR-003 in France, a small organic farm run by farmers Guillaume and Virginie, is another initiative. Each initiative received a code as soon as it was considered as a data source, whether or not it finally became a case.

To sum up: the initiative is the sampling unit, the real-world entity that acts and seeks to change the system. The case is the material collected from it, digitised and archived for research.

The case must help us understand how the system works and evolves. We are interested in entities as parts of a system, so the material collected must document the connections of the initiative to the system and how it is embedded in it.

### 2.1.2 Sample selection

Once the type of entity is chosen, we must decide which units to sample. Sampling a complete subsystem, say all actors of the dairy sector, is impossible: they are too many, from primary producers to retail shops, makers of milking machines and refrigerators, veterinarians and so on. Nor can a food system be sampled representatively, because the relations between agents are part of the system.

We therefore chose to maximise diversity, with the intention of building typologies and ontologies of agents and of their interactions. We cannot reconstruct the system, but we hope to understand the mechanisms of its change. So we sample initiatives that are diverse, typical of their kind, and active in the transition. This is what we ask our local contacts to scout for.

The criteria for 'interesting' initiatives - well-inserted in the system (*embeddedness*: dense, acknowledged ties to surrounding stakeholders and communities), active for several years (*robustness*: long-lived enough for the solidity of the model to be observable), and recognisably attempting to change the system (*vision*: an articulated ambition to improve the local world and the wider ecosystem) - were complemented in the field by two practical filters, *relevance* and *exemplarity*: the case had to promise usable insight and to stand either as typical of a recognisable category or as inspiring in its own right. Some of these criteria are subjective, and they are assessed from what can be known from the case before the interview itself (e.g. from their communication on their website, local contacts' knowledge…). Applied as heuristics rather than thresholds, they may seem vague, but they yielded a satisfactorily diverse sample, even if we have some cases that are redundant, or less "interesting".

Initiatives are first scouted by our local partners who know the sector, by browsing the web with keywords and snowballing, and through relevant lists and databases of organizations (such as local business associations and sustainability prizes). The spotted initiatives are then described with a brief due diligence web search with the description produced by hand or AI, and submitted to the project's research leaders, the PIs, for selection and approval. Because many of the initiatives studied were small food producers and retailers, they often had little online presence. Local contact, and in many cases an informal first visit, was needed to complete the assessment. The process of identifying and selecting is outlined in the Scouting and selection step of the Data Collection Pipeline section below. Not all selected initiatives finally led to cases, subject to availability of access, availability of local interviewers, and consent. The selection ratios were as follows: scouted to chosen, 45% (1,451 of the 3,247 leads logged in the master case log passed validation); chosen to completed case, 50% (731 closed at the 4 July 2026 snapshot); an overall conversion of 22%. The second figure is low because we overshot our target and stopped converting validated initiatives into cases (see Lessons learned): we had validated more initiatives than we could process. The monthly conversion rate from validated to completed rose with experience and was close to 90% when we stopped. Leads still in process make these ratios floors (see Appendix 4).

## 2.2 Data collection

Data collection on site follows the Data Collection Protocol (available from the authors on request). The protocol gives the field investigator all the instructions for collecting the data that build the case, including the topic guide for the interview. Its table of contents is reproduced in Appendix 7, to show what a field-ready LSQR protocol has to cover. The protocol is not used alone. It comes with four training videos, some fifty minutes in all, which walk the interviewer through the same procedures: the project, the logs and the contact phase; preparing the interview; executing the interview; finalising the mission. Interviewers watch them before their first case, and the protocol points to them stage by stage.

Each case contains basic facts describing the data source (location, history, business model…) as is usual in case studies. Because the Socioscope wants to understand not a self-standing case but an element of a larger system, we designed a specific tool, the transaction grid (Lahlou, 2024).

The transaction grid lists the stakeholders with whom the initiative interacts. For each transaction, we record what the initiative gives and gets, from its own perspective and, where possible, as seen by the other party. The grid makes the exploration of relationships systematic. The interviewer lists with the interviewee every stakeholder the initiative transacts with: clients, providers, partners, competitors, local powers, and so on. She then makes them describe each transaction open-endedly. The exchanges go well beyond money: in-kind support, information, access, social capital, legitimacy, ideological satisfaction, regulatory compliance, and so on. This is the classic insight of economic sociology: economic action is embedded in social relations, not conducted between atomised agents (Granovetter, 1985).

An exchange from one of our cases in the United States (US-070) shows how a line of the grid is elicited and filled in the participants' own words:

> *Interviewer: "From your perspective, what do you give them? And then from their perspective, what do they get from you if there is a distinction between those things?"*
>
> *Interviewee: "I think that answer has changed a lot […] there's the logistics side of things of, like, providing the food, providing and helping with refrigeration, helping with, you know, storage and all of that. […] We're providing that to them. And then […] they in turn are giving us […] They're opening their space for us […] to have that food and to be a* *partner of ours and to help us. Like, we can't do the work without them. And so they help accelerate that mission for us."*

The transaction grid rests on the notion of social contract: a structural, multi-transaction agreement of the form "if you give this, you get that" (Lahlou, 2024), sustained over time by reciprocal expectations. The logic is that of gift and counter-gift (Mauss, 1990), not of the spot market. Social contracts use many currencies besides goods and money: information, reputation, access, safety, trust. They are partly informal and open-ended, since social exchange creates "unspecified obligations" (Blau, 1964: chapter 4) that bind the participants and let the contract change over time.

Interestingly, the transaction grid also compensates in part for the limits of sampling. We cannot sample all components of a system, but some appear in the transactions our interviewees describe. For example, in case DK-019 , the founder of the initiative recounts being paid by a Swedish municipality to come and inspire a citizens' meeting on community farming, and by a Copenhagen culinary school to develop teaching material on regenerative ocean gardening. Neither party was interviewed; yet what each sought (inspiration, a working template, teaching material) and what each gave (fees) is recoverable from the interviewee's side of the grid.

Each transaction can appear twice in the grid: as the initiative sees it, and as the initiative believes its partner sees it. Our data show that these are frequently not the same (Lahlou et al., 2024). For example, a social landlord renting space to a women's-employment kitchen is not only collecting rent; it is buying reputation for a deprived neighbourhood. Capturing both perspectives is what brings such hidden "currencies" of exchange to light.

AI models extract the transactions from the verbatim transcripts at scale. Not all elements of the grid come from the moment when the interviewer fills it in with the interviewee; many are collated from other parts of the case. Table 1 gives a raw example of a grid reconstructed by AI analysis of the two interview recordings of a Spanish case; names of persons, places and organisations have been removed.

*Table 1. Transaction grid of a Spanish case (ES-005), coded by AI from the two interview recordings. Names of persons, places, organisations and programmes have been removed or replaced by descriptors in square brackets.*

| Entity / Stakeholder | What the initiative gives | What the initiative gets | What the Other gives | What the Other gets | Quotes |
|---|---|---|---|---|---|
| **Small-scale fishermen (local fishermen's brotherhood; co-founder of the initiative)** | Hope and a sense of doing right by the sea; visibility for a fishery nobody knew existed; scientific knowledge, materials and method (cuttlefish beds, incubators) | The fishermen's voluntary daily labour at sea, and the insider legitimacy without which the project could not exist at all | Deployment of the cuttlefish beds, recovery of eggs from their fishing gear and adherence to sustainable fishing rules, done unpaid for years | Pride and well-being, growing demand for their cuttlefish, and the prospect of a fairer price for their catch | - "It gives them hope, it gives them a sense of well-being that they are doing things right, that they are doing something for the environment. It gives them visibility, because since the cuttlefish project people didn't even know that cuttlefish were fished here, the demand for cuttlefish has increased."<br>- "Without them there is no project, without them there would be no branches. Without them, they are the reason. Yes, they are the raw material."<br>- "They are not obliged to do anything. And they get nothing in return. Well, now I hope it will change, but they don't get anything. They do it completely voluntarily."<br>- "As a biologist, I would never have entered if it wasn't for the fact that I was going hand in hand with a fisherman, with [the co-founder]. "<br>- "What we have to achieve is that their cuttlefish, their squid, are sold more expensively or at a fair price." |

*Table 1. Transaction grid of a Spanish case (ES-005), coded by AI from the two interview recordings. Names of persons, places, organisations and programmes have been removed or replaced by descriptors in square brackets.*

| Entity / Stakeholder | What the initiative gives | What the initiative gets | What the Other gives | What the Other gets | Quotes |
|---|---|---|---|---|---|
| **Other fishermen's associations (two neighbouring ports)** | A tested method and materials, and a shared framework that brought three previously distant brotherhoods to work together | Scale for the pilot in its first years, and a test of whether the model transfers beyond the home port | Participation in the 2016 multi-port pilot; both have since withdrawn, one to run its own version and the other through disengagement | Access to the technique and, for one of them, an approach it now runs independently | - "That year we managed to do it in [the three ports], a large-scale pilot test, well, medium-scale, in which we were testing different techniques, different methods, in different ports to see what was happening."<br>- "I was very proud when we were able to unite the three brotherhoods. At that moment I felt very good, very proud. I was able to bring together three brotherhoods that at first didn't get along very well."<br>- "[One port] dropped out because they decided that they were doing it their own way. [The other port] also left because, as we said before, they are fishermen who don't really want to change things"<br>- "If [that port] is no longer motivated, it is because we have not known how to do it well. So we have to motivate them again." |

*Table 1. Transaction grid of a Spanish case (ES-005), coded by AI from the two interview recordings. Names of persons, places, organisations and programmes have been removed or replaced by descriptors in square brackets.*

| Entity / Stakeholder | What the initiative gives | What the initiative gets | What the Other gives | What the Other gets | Quotes |
|---|---|---|---|---|---|
| **Town council (local municipality)** | Visibility for the council's environmental and sustainability policy, and visible support for small-scale fishing as a local sector | Money and contracted activities that fund outreach to schools and tourists; a hoped-for subsidy for monitoring the cuttlefish beds | Funding and paid commissions for activities, including activities offered free to the public | Policy visibility on sustainability and a way of giving back to a sector it considers important | - "To the city council, well, we bring visibility, we bring visibility in terms of their policies of what do you call this? Of this very fashionable word that is now environmental and durability, of all this that is now fashionable."<br>- "Money that allows us to do activities, impact schools and tourists."<br>- "It is in their interest to promote small-scale fishing, because it is an important sector."<br>- "We are supporting small-scale fishing and by supporting the cuttlefish project, they are giving back to the fishermen."<br>- "Sometimes they are free activities that are paid for by the municipality."<br>- "And in this way what we also want, apart from that, is that the town [councils] help us to subsidise the monitoring of the sepieras." |
| **Natural Park (adjacent marine and coastal park)** | The marine component of the park's school programme, filling an age bracket its own offer does not cover | Paid school activities that provide reliable income and access to schoolchildren it could not otherwise reach | Payment for the school sessions, a designated space and permit for the cuttlefish beds, and early research collaboration on egg-laying substrates | A ready-made solution for its conservation and dissemination training programme, [park-in-schools] | - "The schools in the municipalities in the park pass through here and the park pays us for these activities which are free for the schools. So it benefits us a lot because, well, because it's an income that's very good, it's a way of accessing these schoolchildren that |

*Table 1. Transaction grid of a Spanish case (ES-005), coded by AI from the two interview recordings. Names of persons, places, organisations and programmes have been removed or replaced by descriptors in square brackets.*

| Entity / Stakeholder | What the initiative gives | What the initiative gets | What the Other gives | What the Other gets | Quotes |
|---|---|---|---|---|---|
| | | | | | otherwise we wouldn't be able to reach them."<br>- "It's part of their training programme for the conservation and dissemination of the park. Well, we fit them in very well, we solve a problem for them at certain ages."<br>- "They have a project called [park-in-schools].<br>So [the project] provides a solution to this issue for children"<br>- "I was working with the natural park, I am a biologist, I was working with the natural park doing some experiments with cuttlefish branches, pine branches, intiscla branches." |
| **Fisheries administration (regional fisheries department, species co-management body, local fisheries action group)** | Pressure and a working example that forced the administration to engage; the co-management body itself grew out of the project | A permit and a designated marine space for the cuttlefish beds, and a route to public fisheries funding | Authorisation and allocated marine space; a funding channel through the local fisheries action group | A demonstrable, locally-owned sustainable fishing initiative it can point to, and continued prompting to act | - "This is where the co-management body came out of, and the administration got involved, which is very important."<br>- "It's very good to have made the administration have to put their feet in the water and get wet and get involved, that's very good."<br>- "You are there as we are the stone in shoes that we are motivating, this is very good."<br>- "What they have done, they have given us a space |

*Table 1. Transaction grid of a Spanish case (ES-005), coded by AI from the two interview recordings. Names of persons, places, organisations and programmes have been removed or replaced by descriptors in square brackets.*

| Entity / Stakeholder | What the initiative gives | What the initiative gets | What the Other gives | What the Other gets | Quotes |
|---|---|---|---|---|---|
| | | | | | and we can put the epieras there. That is now in place. So, at that level, we are very happy, obviously, that we can put the epieras without having to ask for permission"<br>- "The fishermen are more affected by the fisheries department. But we have very little to do with it. Very little relationship."<br>- "From [the local fisheries action group].<br>" |
| **Tourists and visitors** | A tourist product available nowhere else: beach talks, snorkelling to a visitable replica cuttlefish bed, and boat trips explaining the project | Emotional feedback and gratitude, paid income where the activity is bundled with a boat trip, and an opening to change consumption habits | Attention and money spent locally, and willingness to pay when the activity is packaged with a boat trip | A differentiated experience and a first encounter with the state of the sea and the world of cephalopods | - "We are giving them a different activity, that is, a differentiated tourist product, a different activity that cannot be found anywhere else."<br>- "We put a replica of the cuttlefish at a depth of 2 m on the beach, always within the marked area, so that the boats don't run over people. And then what we do is that we give a talk on the beach, we show them the eggs, then we go swimming, snorkelling and we go to visit the visitable cuttlefish"<br>- "Basically we get the emotional feedback of gratitude after doing this activity. We have tried to do the activities for a fee, but it's difficult" |

*Table 1. Transaction grid of a Spanish case (ES-005), coded by AI from the two interview recordings. Names of persons, places, organisations and programmes have been removed or replaced by descriptors in square brackets.*

| Entity / Stakeholder | What the initiative gives | What the initiative gets | What the Other gives | What the Other gets | Quotes |
|---|---|---|---|---|---|
| | | | | | - "A traditional boat that takes tourists, we do a trip and we explain [the project] , we explain the environment, the fishermen, that works well. People are willing to pay for that."<br>- "They have brought up things that we had no idea about, for example, that the sea is very bad, that people don't even think about it" |
| **Consumers (through a direct-sale channel, currently paused)** | Traceable, top-quality local cuttlefish, and the knowledge of who caught it and what that fisherman does for the sea | Money that flows back to fishermen taking direct environmental action - the mechanism that closes the circle | Willingness to pay a fair price for a local, traceable product instead of an anonymous industrial one | A product of the highest quality and the well-being of knowing their money is funding marine improvement | - "My intention is that through [the project], you can eat a cuttlefish and know that [the fisherman] has caught it. And that you know that [the fisherman] is taking direct action to improve the cuttlefish, the environment where the cuttlefish live. And that your money is paying [the fisherman] to carry out these actions. That is the objective."<br>- "They are receiving a product of the highest quality, they know that they are also subsidising a project to improve the environment, so they feel good. Knowing that your money is going towards something positive is a feeling of well-being."<br>- "As the [direct-sale] link has been broken , because now there |

*Table 1. Transaction grid of a Spanish case (ES-005), coded by AI from the two interview recordings. Names of persons, places, organisations and programmes have been removed or replaced by descriptors in square brackets.*

| Entity / Stakeholder | What the initiative gives | What the initiative gets | What the Other gives | What the Other gets | Quotes |
|---|---|---|---|---|---|
| | | | | | is no consumer"<br>- "We link it to the fact that if we consume local products we are paying, our money goes to local fishermen who can take direct action to improve the situation." |
| **Schools and schoolchildren (school programme)** | Environmental education sessions and hands-on activities, including releasing a named baby cuttlefish | Impact on future consumers and possible future fishermen, and paid access via the park and the council | Access to their pupils and classroom time, mediated by the park and the town council | A marine education offer that changes how children see the sea | - "We have created a project called [name], we are very original, just to have an impact on the school population and this year we want to promote it to reach future consumers or who knows, future fishermen, who knows."<br>- "One of the activities we do with the children is that we have a baby cuttlefish and then a glass and they release it. The emotional moment of releasing a cuttlefish and they give it a name and they are releasing it, that's very nice."<br>- "There are many children that come to me and no, no, the beach is for bathing and that's it. And they come with this mentality. When the talk is over they look at it in a different way."<br>- "This year we have joined forces with a foundation and a |

*Table 1. Transaction grid of a Spanish case (ES-005), coded by AI from the two interview recordings. Names of persons, places, organisations and programmes have been removed or replaced by descriptors in square brackets.*

| Entity / Stakeholder | What the initiative gives | What the initiative gets | What the Other gives | What the Other gets | Quotes |
|---|---|---|---|---|---|
| | | | | | company to promote this product and reach the schools." |
| **Local businesses and private sponsors (mechanics, flats, campsites, yacht clubs; a foundation and an educational services company)** | A local project they can believe in and be part of, rooting environmental action in the local economy rather than in a vertical grant | Private money that makes the funding hybrid, and specialised educational delivery capacity | Cash contributions raised door to door, and now professional educational services for the schools programme | Participation and ownership in a visible local initiative | - "One thing I like to explain a lot is that the resources we have had have been very hybrid. What does that mean? A mixture of public money and private money. This seems to me to be very important"<br>- "We could have gone to Europe to ask for a project, they would have given it to us, they would have given us the money vertically, we would have opened the project and it would have ended. And that would have been a failure for our taste. What we wanted was for this to enter society and the best way to enter society is to involve local companies to participate in the project."<br>- "We are playing salesmen, a fisherman and a biologist playing door to door salesmen and we got mechanics, flats, |

*Table 1. Transaction grid of a Spanish case (ES-005), coded by AI from the two interview recordings. Names of persons, places, organisations and programmes have been removed or replaced by descriptors in square brackets.*

| Entity / Stakeholder | What the initiative gives | What the initiative gets | What the Other gives | What the Other gets | Quotes |
|---|---|---|---|---|---|
| | | | | | campsites, yacht clubs, all those who believed in this project and put money aside"<br>- "In that year, there was no tourism here. The companies said no, I'm sorry, but I can't give you any more money. Many participants dropped out." |
| **Divers and diving centres (local divers' association; the area is a diving hotspot)** | A purpose and a hook for their dives - seahorses, cephalopods and an explanation of the project - plus training through underwater photography | Citizen-science data, a photographic record of marine biodiversity, and a replicable pilot for the wider Mediterranean | Photos and images of the species they observe, and monitoring effort on the cuttlefish beds | A differentiated diving experience and knowledge of the marine environment they dive in | - "We also want to reach out to [divers] here in [the town] where we are a hotspot for diving. So this year we are trying to reach out to [divers] to educate them through photography, to take them to see the seahorses, to take them to take pictures with the hook"<br>- "I want [divers] to help us through their photos and images, to make a compilation of the biodiversity that exists, the invasive or non-invasive species that exist, looking for the hook of seahorses"<br>- "Through citizen science I want to collect data and be able to see what is going on."<br>- "With the [divers'] association.<br>Here we want to mix that, the fishermen, the divers and the park, to make a trinomial that I think |

*Table 1. Transaction grid of a Spanish case (ES-005), coded by AI from the two interview recordings. Names of persons, places, organisations and programmes have been removed or replaced by descriptors in square brackets.*

| Entity / Stakeholder | What the initiative gives | What the initiative gets | What the Other gives | What the Other gets | Quotes |
|---|---|---|---|---|---|
| | | | | | can give very interesting results."<br>- "If we manage to replicate it through the divers, this will make it much easier to disperse it throughout the Mediterranean." |

We are still exploring techniques to fill in transaction grids in the most detailed and stable way. Extraction is sensitive to the models and prompts, but also to the extraction strategy and to its granularity (components of a transaction, or full descriptions). Named-entity recognition is still perfectible. Nevertheless it is already clear that the data collected yield substantial transaction grids, although they will always remain incomplete.

Sampling also snowballs along the transaction grid: when a case names a key stakeholder, that stakeholder may itself become a case. Each initiative is thus a window onto the wider network. This link-tracing logic comes from chain-referral sampling (Heckathorn, 1997), but we use it to follow the structure of a system, not to reach a hidden population. We applied it deliberately in our “zones of interest” (the municipality of Aarhus (Denmark); the Dijon metropolis (France); the Breede Valley in South Africa; and a Colombian value chain), where we densified the sample to get a more complete view of the system.

### 2.3 The Corpus and the Gate

The Corpus is a digital repository of all the cases.

One principle is that the Corpus can only grow: nothing is erased. This is a safeguard. When a team handles files, uncertainties and errors are inevitable. No erasure is the surest way to keep out of such problems.

The Corpus is structured by cases, each of them has a code number. For example, Case AR-068 contains, as of July 2026, 41 files: 7 transcripts of interview recordings, 34 original media files including interviews, debrief and supplementary material, for a total of some 7.9 GB. The transcripts are translated from the original language (Spanish) to English (so we have the original video or sound file, the original-language transcript, and its English translation). The supplementary material includes pack shots of product, the facility etc. The material usually includes a video tour of the facility.

Each file name states explicitly what type of data the file contains, and is paired with a comment describing what it is and its successive transformations. Every processed file (e.g., files we derived from original uploads, such as transcripts from interview videos) also contains its own log file, with details such as what version of what AI model was used in transcription and translation.

The Corpus is accessible through the Gate, an interface which documents the operations made on the Corpus. The Gate restricts and logs access to the Corpus. As said above, one of the principles of the Corpus is that it can only increase. When an original file is re-processed, e.g. transcribed with a different model, the new transcript is simply added to the case, and this is noted on a ledger in the Gate.

These precautions may look excessive. But such a Corpus costs millions to build, its value grows as it grows, and it would be foolish to waste it by poor management.

## 2.4 The social contract

Collecting deep data depends on the active collaboration of busy people at the sources. 'Interesting' cases often raise public interest, and may lead to research fatigue: scientists and journalists arrive, take time, ask intrusive questions that may touch trade secrets or failures, and leave little behind (Clark, 2008). The Socioscope is wary of adding to this fatigue. It treats participation as a balanced "social contract" (Lahlou, 2024), not as data extraction: we ask what is in it for the initiative, and answer with real returns. A social contract is the trade of a role (what others can expect from you) for a status (what you can expect from others). We aim to set balanced social contracts with our data sources.

In Phase 1, in exchange for their time and access, participating interviewees received a professional, short, promotional video (around four minutes) in which they present their initiative. They also got a dedicated page on the Socioscope website (for which they can suggest edits), and, ideally, membership of a global community of like-minded initiatives. Both are optional, and some (few, especially official bodies such as municipalities) opted out. They also gain the recognition of being studied by prestigious research institutions, and the experience of reflecting on their own practice with experts during the interview. We also sometimes send them a certificate of participation which they can frame and display.

In return, the research team obtained the in-depth interview, a filmed tour of the facility, access to confidential business information, sustainability reports and supplementary materials, and the signed informed consent form and image-rights authorisations. Rewarding data providers for their participation has always been a concern for social scientists (e.g., Mauss, 1990; Clark, 2008; Maiter et al., 2008; Singer & Ye, 2013) and some practitioners form local communities of knowledge and practice (Lave & Wenger, 1991). The Socioscope website is an attempt to build an online community that gives initiatives the visibility they look for, and to maintain a relationship that may make it easier to collect more data from them later.

# 3 Data enrichment and release for analysis

Collection and curation do not end when analysis begins. Analysis often reveals that the Corpus needs further treatment, and additional data. The pipeline is therefore a cycle: collect, curate, analyse, find what is missing, enrich, analyse again. This section describes the layer that supports this abductive movement between data and interpretation (Timmermans & Tavory, 2012).

## 3.1 Corpus design

Data remain usable only as long as the institutions and practices that curate them. The cost of keeping data usable is chronically underestimated (Borgman, 2015), and the maintenance budget often non-existent. The e-science literature treats provenance (the derivation history of a data product from its original sources) as what makes computational results trustworthy, auditable and reproducible (Simmhan, Plale, & Gannon, 2005). This literature has largely ignored qualitative research, not because it does not apply, but because hand-made corpora were small enough to live in one researcher's memory and filing cabinet. At the Socioscope's scale this is no longer possible: interviews need versioned releases, audit trails and provenance tracking, as does the process itself.

The transcribed / standardised text never replaces the source. Links back to the original audio and video are kept, so that an analyst can always return to the spoken words, check the context and look for translation loss. The recording becomes a structured, queryable object; the original is never overwritten. We log the series of operations performed (which LLM, which version, and so on), because tools improve and a later pass may give better results than the first, as happened with our utterance tagging.

Once we have uploaded a file, the Gate curates it into the Corpus. It pseudonymises the text where required and keeps a secure mapping from name to pseudonym; it generates metadata and a machine-readable audit record; it runs a content quality check that flags issues for human attention; and it commits a numbered version of the case with a log entry. The first commit, v0.0.1, marks the case Quality Confirmed. A ledger sits alongside the raw and transformed files, tracking which version is current.

The composition of the case folder (collected material, metadata and logs, and analysis products) is detailed in Appendix 3 (supplementary material).

## 3.2 Preserving provenance

One of the coming problems of research in the era of AI is fake data. The provenance of the data will become a key component of research quality, and data control an essential part of scientific work and its validation. That is why we keep evidence of what we call, by analogy with museums, “data provenance”: where, when, and how the data were harvested and curated, so that any extract can carry a label. Here is an example of a labelled speech turn (this one is short; some run much longer):

CH003_0000149, CH-003-2024-01-01-INTERVIEW_AUDIO-AUDIO-1-0.1.0-EU-FR-EN_deepl.csv, 12, SPEAKER_01, Olivia B., interviewee, CH-003-2024-01-01-INTERVIEW_AUDIO-AUDIO-1-0.0.1-EU-FR-EN.m4a | Olivia B., 71.405, 84.1, 00:01:11.405, 00:01:24.100, fr,

*"On entretient l'ensemble du quartier des vergers, donc c'est un gros quartier de à peu près 3000 habitants sur à peu près 16 hectares, on a 16 hectares en entretien."*

*"We maintain the entire Vergers neighbourhood, which is a large neighbourhood with about 3,000 residents spread over roughly 16 hectares."*

(The capitalised "Vergers" was recognised as a proper name during transcription.)

Our approach is radical, and costly in disk space and tracing effort: originals are immutable. We never destroy data, we only add to them. Every enrichment (transcript, translation, tag, extracted grid cell, scraped supplementary material) is a traced layer on top of the original, with the chain back to the source preserved, in keeping with the FAIR principles for scientific data management (Wilkinson et al., 2016). We also assign a SID (Socioscope ID) to every speech turn, shown as CH003_0000149 in the example above, so that findings and analyses can be traced back to the verbatim. During analysis we often need to read, or listen to, a larger context, for example to check an interpretation by a human analyst or by an AI prompt.

The links back to the source (see Corpus design) are used more than expected: because nuances are better understood in the original language, and because the original material reveals errors made along the chain, for instance speech turns misattributed when interviewer and interviewee have similar voices. Attributing an interviewer’s words to the interviewee would be a real problem for analysis; the link to the source audio is what lets us catch it. These errors get rarer as transcription tools improve, but they still happen.

Tracing matters for two reasons. First, integrity. Tools change much faster than the data (we change LLMs and their versions regularly), so the project must be able to re-run any step on the untouched originals with better tools. This is only safe if originals are never overwritten and every transformation is recorded. Second, traceability. Each enrichment carries its own provenance, so the Corpus accumulates value, not just files. Having the data is not enough; we must know when and how each piece was acquired and modified. Organisations change, and a consent form may turn out to have been signed by someone who has since left. Comparability and data quality must therefore be actively maintained; every point of access or treatment can introduce error. The same discipline applies to the project's own records; the resulting documentation library is described with the pipeline's outputs below.

### 3.3 Correcting upstream and releasing for analysis

The chain is not purely linear. Quality control sometimes finds something that requires going back upstream to complete or correct: a misspelled proper name, or a data point never properly recorded. The pipeline is better described as a linear stream with quality control at each step, several correction loops, and an overall layer of access control and data management. The detailed pipeline (the numbered events with, for each, the actor, the action performed and the file output) is provided as Appendix 2 (supplementary material).

For analysis, the Gate delivers a scoped, filtered retrieval on request rather than raw files. It can freeze an immutable, citable snapshot of the Corpus, so that an analysis rests on a fixed, reproducible state. It controls access and exports, and every extraction is logged against a whitelist. This is where the pipeline connects to analysis and where the loop begins: analysis routinely sends cases back up the chain for enrichment that becomes necessary only once interpretation is under way.

### 3.4 The Socioscope Gate, Software Development Kit, SDK: tracing research inputs and prompts

Provenance discipline (see Preserving provenance) is only as strong as its weakest link, and once a snapshot is handed over for analysis the link is easy to lose. Left to themselves, researchers running an LLM over the data save the result wherever they happen to, with no durable record of which prompt, which model version or which subset of cases produced it. Provenance then lives in the researcher's head, notebook or chat thread: recoverable today, gone when someone asks six months later what produced a finding.

We first dealt with this through the dashboard, now we do it with the Gate which logs everything. We are exploring developing a specific SDK

# 4 The Socioscope installation: equipment, personnel and processes

## 4.1 Equipment

The equipment layer covers three needs: what the field and the central team need to collect the data; what the data need to be stored, processed and curated; and what the social contract needs to be fulfilled.

Field equipment. Each interviewer operates a standardised audiovisual kit. In phase 1: the DJI Pocket 3 Creator Combo, a stabilised 4K pocket camera with gimbal, wireless microphone, tripod and spare batteries, complemented by the interviewer's own laptop and smartphone, an audio recorder for the interview, and a notepad for jotting points and letting interviewees sketch. The pre-visit checklist of the protocol makes equipment checks an explicit step of every visit: batteries and spares, microphone, a recording test, operation with and without internet, a backup recording device. A single consumer-grade kit keeps image and sound quality homogeneous across 37 countries, makes training and troubleshooting generic, and keeps the unit cost low enough to absorb shipping, insurance and occasional loss.

**Equipment logistics.** Kits are dispatched to interviewers once a case is confirmed, and readiness is checked before the visit (see Data Collection Pipeline); where a regional coordinator is in place, kits circulate locally between interviewers rather than transiting through the central team.

**Communication**. Every member of the field team receives a project e-mail address (@thesocioscope.org). Field coordination runs on ordinary channels (e-mail, telephone, messaging), backed during the collection period by a permanent help desk and hotline. The central dashboard (described below) gives the whole field team a shared, continuously updated view of every case.

**Headquarters**. The field is managed by a central field team and intermediaries in local geographies, e.g. consulting firms or universities that act as regional proxies, so that interviewers are managed under local rules. All this requires the usual infrastructure of an efficient organisation: offices, workstations, legal and administrative cover. These are not details. The legitimacy of the project in the eyes of local initiatives depends on the quality of that organisation, because that is what they check before deciding to participate: the project needs a website, field team members need e-mail addresses in the project's name, business cards, and so on. Administrative logistics are not a detail either: hundreds of cases mean hundreds of payments in different currencies, and a control system for execution and quality, with staff and procedures (who answers an interviewer who says she did not receive her transfer?).

**Social-contract logistics.** Fulfilling the social contract has its own small supply chain: leaflets presenting the project in the local language, which interviewers leave with initiatives; the goodies package and the printed participation certificate sent after the visit; and the worldwide shipping of these items by the Socioscope team. They cost little, but they need the same logistical care as the recording equipment: a certificate that never arrives quietly damages the relationship the social contract is meant to sustain.

**Premises**. The core Socioscope team members are based at the Paris Institute for Advanced Study, which hosts the project administratively, and the Vienna Complexity Science Hub. The central team that oversees data collection has an open space and neighbouring offices in the premises of the Paris IAS. This proximity matters for handling the technical and logistical problems of data collection.

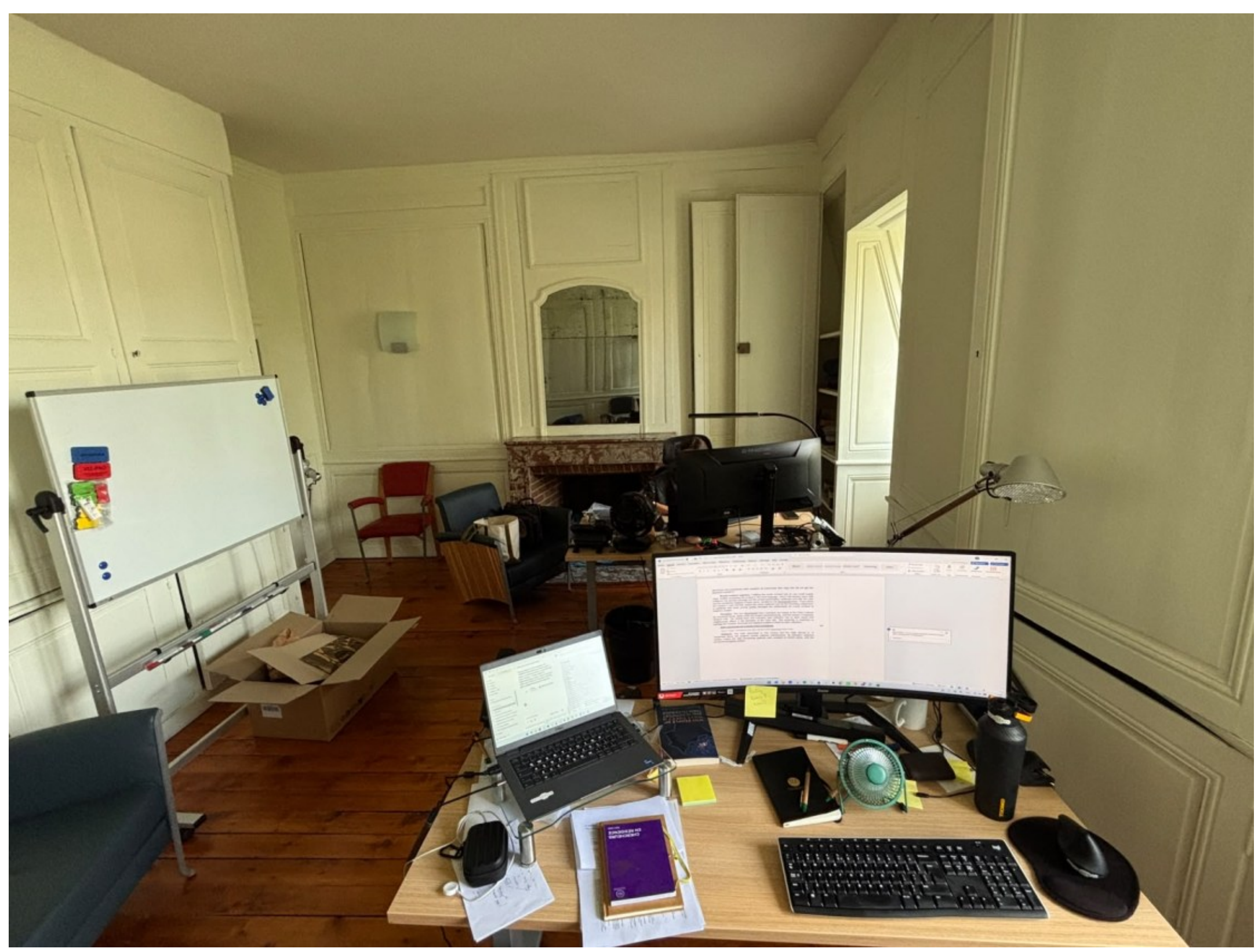

*Figure 1. Workstations of the central data-collection team in Paris.*

**Software.** The Gate (described in The Corpus and the Gate above) is, in equipment terms, the project's central software infrastructure. Its modules: selective access control, the data processing pipeline, web scrapers to enrich cases, and the provenance/logging system.

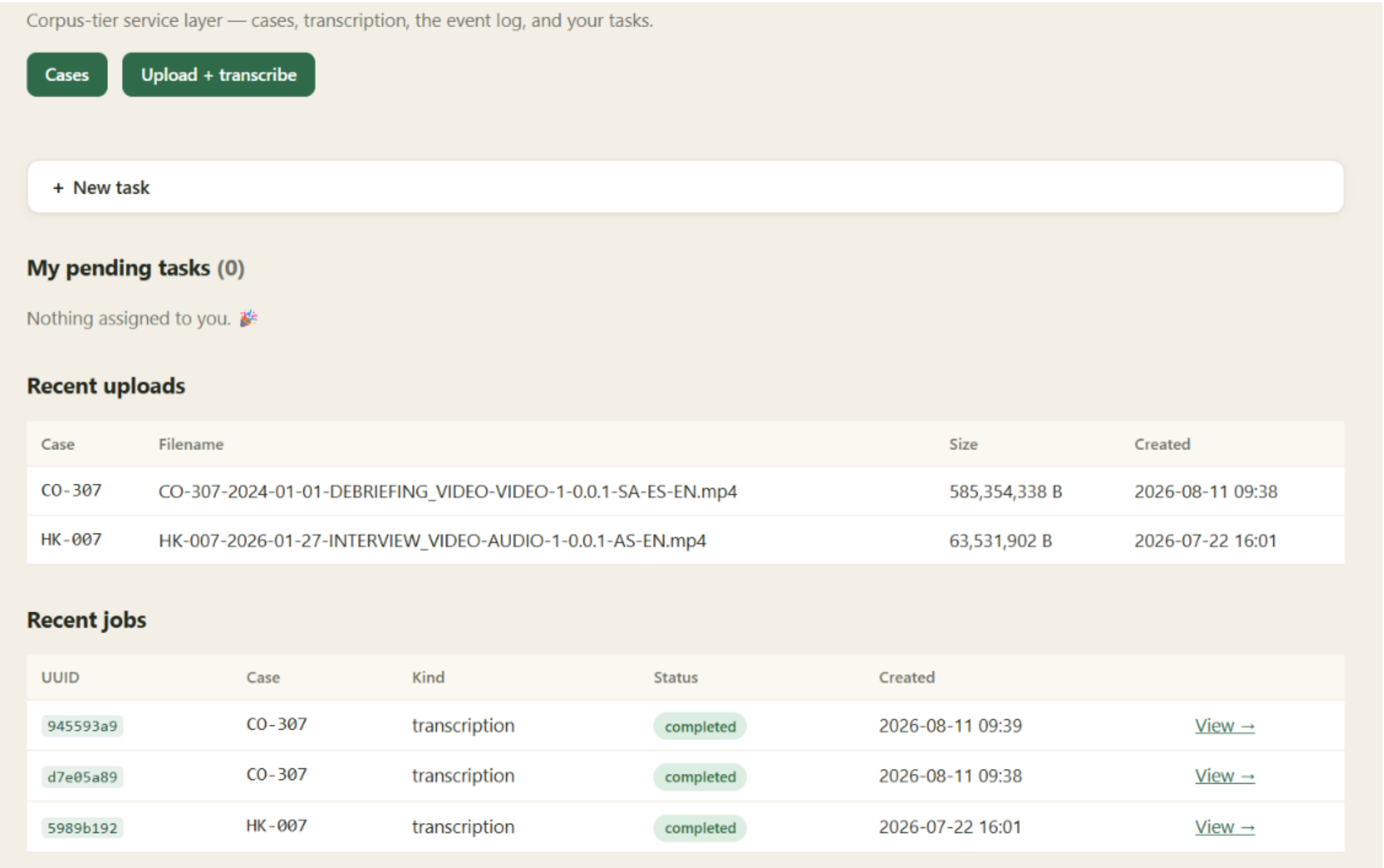

*Figure 2. Gate upload landing interface.*

***Upload, storage and Gate infrastructure.*** Collected material is uploaded from the field through the Gate, which authorises the upload, renames and provenance-stamps the files, and writes the audit log. Behind the Gate, the Corpus resides on redundant cloud object storage, with a database serving as index.

**The project's working documents** are hosted in a separate cloud bucket (accessible via a separate internal website; see *The Internal Website* below), and the public website is a third, strictly separated block (see *The Public Website* below). The proof of concept, run on consumer cloud drives, showed their limits for interaction, access control and multi-gigabyte video files; that experience led to the development of the Gate.

***The Internal Website.*** This manages all internal documents, processes, tasks, and resources. The documents and procedures needed to manage the project are so many that an internal website had to be set up to curate and retrieve them. Because field personnel turn over, these documents are used and shared constantly, especially during contracting and training. As of July 2026, this amounted to 1,432 unique documents in fourteen categories, from contracts and invoices to protocols and training materials, plus 93 transcribed recordings of meetings and training sessions; the documentation library is described among the pipeline's outputs below.

***The Public Website.*** This shares updates about the Socioscope project, and connects all initiatives within a mutually beneficial online community: www.thesocioscope.org

The site presents the project's aims and its three strands (data, technology, analysis), headline figures (documented initiatives, countries, interviewers, languages), a world map of the case studies, supporting institutions and the Socioscope team. It also serves the social contract: participating initiatives gain public visibility and can follow the Socioscope project's progress.

Most important, it has a search engine and filters to find initiatives. Users can also connect with initiatives of interest and receive alerts when a new initiative matching their criteria is entered.

***The Central Dashboard.*** This workbook followed data collection step by step during Phase 1 and gave a continuously updated view of each geography. It has now been replaced by the Gate.

It was a shared workbook with one sheet per geography (54 country sheets, 3,247 dated lead rows since 2023), one row per lead, tracking each case from scouting through validation, fieldwork and closure: status, dates, actors, payment validation. Aggregate KPI sheets summarise progress per country; the pipeline metrics in Appendix 4 are extracted from this instrument.

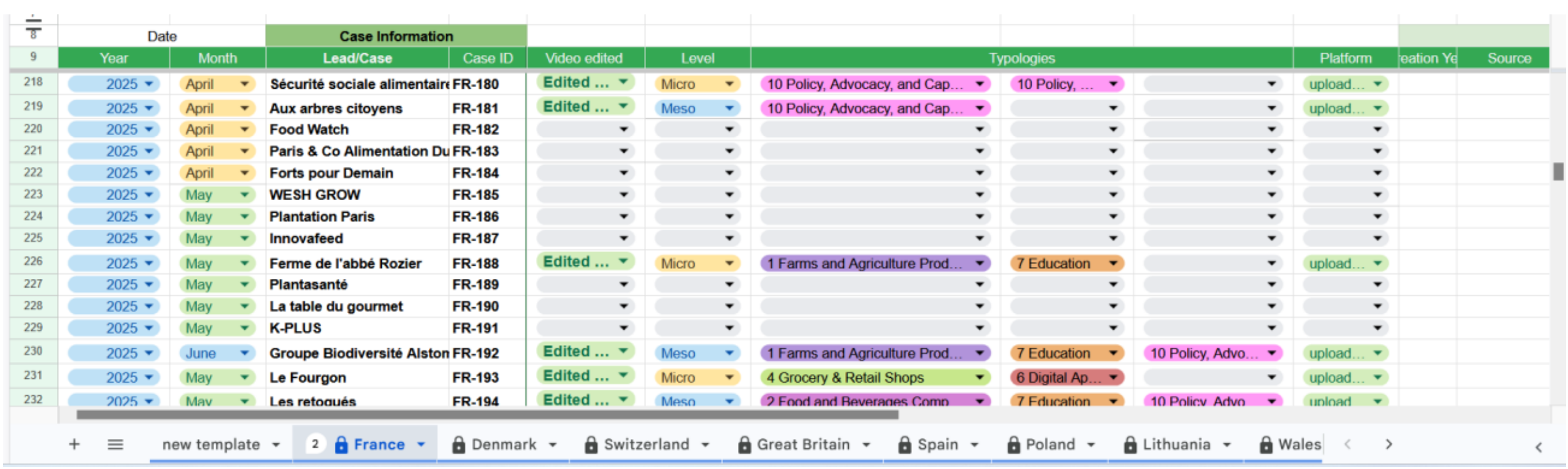

| 8 | Date | | Case Information | | | | | | | | | |
|---|---|---|---|---|---|---|---|---|---|---|---|---|
| 9 | Year | Month | Lead/Case | Case ID | Video edited | Level | Typologies | | | Platform | eation Ye | Source |
| 218 | 2025 | April | Sécurité sociale alimentaire | FR-180 | Edited ... | Micro | 10 Policy, Advocacy, and Cap... | 10 Policy, ... | | upload... | | |
| 219 | 2025 | April | Aux arbres citoyens | FR-181 | Edited ... | Meso | 10 Policy, Advocacy, and Cap... | | | upload... | | |
| 220 | 2025 | April | Food Watch | FR-182 | | | | | | | | |
| 221 | 2025 | April | Paris & Co Alimentation Du | FR-183 | | | | | | | | |
| 222 | 2025 | April | Forts pour Demain | FR-184 | | | | | | | | |
| 223 | 2025 | May | WESH GROW | FR-185 | | | | | | | | |
| 224 | 2025 | May | Plantation Paris | FR-186 | | | | | | | | |
| 225 | 2025 | May | Innovafeed | FR-187 | | | | | | | | |
| 226 | 2025 | May | Ferme de l'abbé Rozier | FR-188 | Edited ... | Micro | 1 Farms and Agriculture Prod... | 7 Education | | upload... | | |
| 227 | 2025 | May | Plantasanté | FR-189 | | | | | | | | |
| 228 | 2025 | May | La table du gourmet | FR-190 | | | | | | | | |
| 229 | 2025 | May | K-PLUS | FR-191 | | | | | | | | |
| 230 | 2025 | June | Groupe Biodiversité Alston | FR-192 | Edited ... | Meso | 1 Farms and Agriculture Prod... | 7 Education | 10 Policy, Advo... | upload... | | |
| 231 | 2025 | May | Le Fourgon | FR-193 | Edited ... | Micro | 4 Grocery & Retail Shops | 6 Digital Ap... | | upload... | | |
| 232 | 2025 | May | Les retoqués | FR-194 | Edited ... | Meso | 2 Food and Beverages Comp | 7 Education | 10 Policy, Advo | upload | | |

new template | 2 France | Denmark | Switzerland | Great Britain | Spain | Poland | Lithuania | Wales

*Figure 3. Extract of the central dashboard.*

## 4.2 Personnel

### 4.2.1 Hiring: data management and interviewers

The Socioscope needed many people in the data collection phase: some 80 interviewers in the field, regional managers and gatekeepers, a core coordination group across Paris and Vienna, three central field managers constantly in touch with regional coordinators and field interviewers, audiovisual staff, computer scientists, subcontractors, with 2 full-time data-collection members (including 1 with legal training) and a quality-control officer.

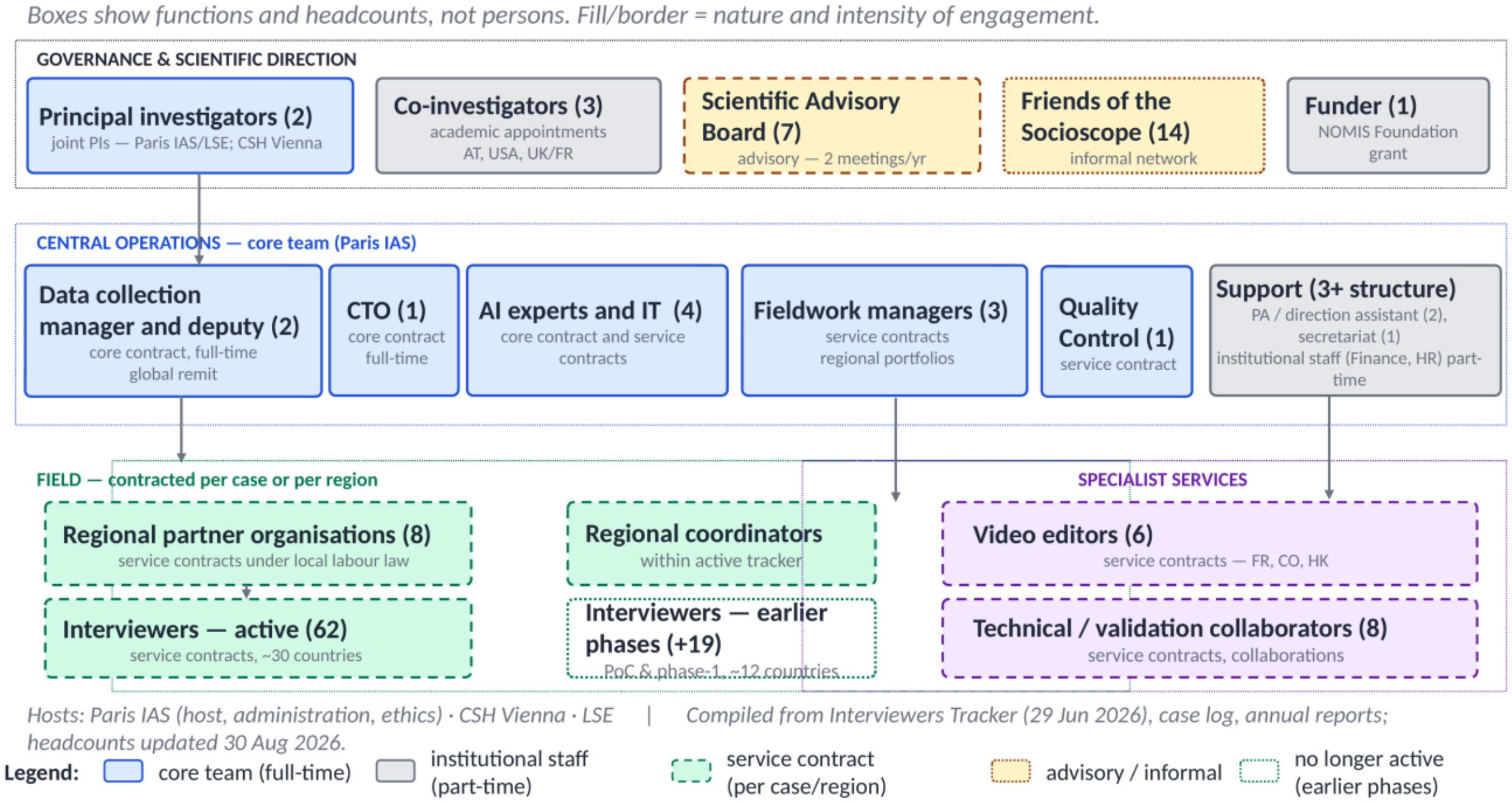

*Figure 4. Structure of the Socioscope project team for data collection.*

### 4.2.2 Training for scale and diversity

Training material is standardised: the Data Collection Protocol and its four videos are the same everywhere. Local interviewers are recruited and trained by qualified staff and work following the same protocol everywhere. This may require flying co-PIs or the data collection manager on site for training sessions, so that depth is achieved locally within a common framework. Regional coordinators in each area (for example Latin America and West Africa) anchor the work in local realities and follow closely the local administration of the project (labour law, insurance, tax). They were also responsible for the equipment issued to their interviewers.

Video editing is run as a separate strand, controlled by the central team, staffed by editors preferably fluent in the language of each interview.

During the proof of concept we found, to our surprise, that even doctoral students and post-docs who claimed dozens of interviews often did not apply properly the interview guidelines. Some seemed badly trained and used the topic guide as a questionnaire, while others tweaked the collection of data towards their own interest. We found that many social scientists are not trained for professional collaborative fieldwork. This is not a trivial detail: only 5 or 6 of the 30 interviewers tested in the proof of concept were reliable enough in teamwork and in following instructions to be kept. Social scientists seem currently trained to work alone, or as a cottage industry. LSQR will require a different type of habitus.

Multimedia capture requires good equipment, skills and organisation. This led us to be explicit down to the detail, and to set up serious training and quality control, with the running costs, time, delays and investments in form that come with them. LSQR is no cottage industry. If the social sciences want to exploit the opportunities offered by AI, they must consider what this implies for the organisation of labs and departments, training, careers and funding. We are moving closer to the way research operates in the natural sciences, typically with larger teams, technicians, sizeable infrastructure, and longer-term projects.

### 4.3 Processes

Collecting comparable deep data across 37 countries, several languages and different kinds of initiatives means that comparability cannot be improvised. It must be engineered through an exacting, controlled and traceable process, and above all through heavy "investments in form" (Thévenot, 1984): rules, protocols, templates, contracts and standard formats. Such standardised formats act as boundary objects that hold their meaning across very different local sites while still allowing each to work in its own way (Star & Griesemer, 1989). These investments slowed the first year of collection, but they made the data combinable. We used a previously tested data-collection protocol for interviewers (a manual of fifty-plus pages, available from the authors on request), designed four training videos (available online at the Socioscope website) and established additional support through a permanent help desk and hotline.

Processes are the third layer of the installation, after equipment and people: the rules, protocols and workflows that turn them into a reproducible pipeline. Processes fall into five families: data-collection processes (the Data Collection Protocol, the topic guide, the pre-visit checklists and the debrief procedure); contractual processes (interviewer service contracts, the two consent forms, image-rights authorisations, and subcontracts with the regional partner organisations); data-management processes (file naming, upload, transcription and translation, quality control and corpus curation, all run through the Gate); management processes (case tracking on the central dashboard, payment clearance conditioned on quality control, budgeting and reporting); and compliance processes (ethics and data protection, described below). Each process exists as a written instrument, kept and versioned in the documentation library on the internal website, so that it can be audited, taught to newcomers and revised with experience.

### 4.3.1 Ethical oversight

All research activities must comply with ethical principles and legal obligations, in particular the EU General Data Protection Regulation (Regulation (EU) 2016/679) on privacy and participant rights. Alongside the operational management, an ethics committee supervises ethical standards. For the Socioscope project, the Data Management Plan and compliance documentation were submitted to the Ethics and Deontology Board of the Paris Institute for Advanced Study (Comité d'Éthique et de Déontologie, CED) first on 29 March 2024, and approved at second examination on 22 January 2025. In parallel, we engaged a law firm specialising in data protection, to conduct a comprehensive GDPR compliance assessment. This engagement included:

- Diagnosis and evaluation of current data processing activities;
- Definition of a compliance action plan;
- Implementation of corrective measures where necessary;
- Production of accountability documentation;
- Recommendations for ongoing risk management and data protection practices.

This phase should not be underestimated in time or cost: compliance took nine months and about €15,000 to be validated by the law firm. Its most consequential output was the retention period. The standard for research data is three to five years, after which the data must be destroyed. That is incompatible with large-scale qualitative research, where the corpus is the durable asset and the point of the pipeline is that data paid for once remain analysable as tools change. We obtained an initial retention period of fifteen years, appropriate for longitudinal research. Anyone attempting a comparable project should plan this negotiation from the outset: a default retention period would silently cap the useful life of the corpus.

Partner research institutions also need their own ethical clearance, and the retention period granted to the project must be reconciled with their local regulations. We partnered with the University of the Western Cape (UWC) and its Centre of Excellence in Food Security, the Tecnológico de Monterrey (TEC) and Aarhus Business School; each had to obtain its own clearance and comply with local rules, which took additional time and resources.

After the initial set-up, an LSQR project should monitor emerging legal and technical issues and keep up with GDPR practice: periodic reviews of consent procedures, retention policies and repository security, and staff training in data protection.

### 4.3.2 Supporting operations: legal and financial requirements

The Socioscope also required operations support: legal compliance, human resources, and finance. This includes compliance with regional laws, region-specific insurance coverage, appropriate anonymisation of data, drafting and maintaining employment contracts, and ensuring that data-collection payments are correctly processed around the world. We chose to rely on local proxies and subcontracting, so as not to overload the central team with local labour law, insurance and hiring. Local contacts also know the context better, and working with them builds local expertise and networks for future collaboration. Supporting functions are an integral part of the process, but their detailed description is beyond the scope of this paper.

# 5 Data journey step by step

The sections above describe the foundations of the pipeline; this section follows it as a sequence of operations. Appendix 2 gives more detail.

## 5.1 Overview

The following figure summarises the main path of the data through the pipeline. As described above, quality control runs at every step, with loops when correction is needed.

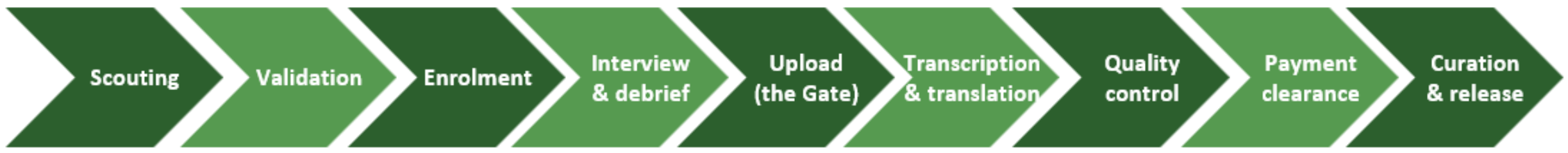

*Figure 5. The main path of data through the Socioscope pipeline.*

## 5.2 Scouting and selection

A case begins as a candidate initiative identified by the Socioscope team and its local contacts (most of them potential interviewers), on the selection criteria given above. The initiative is entered as a prospective case in the Gate, with basic information such as initiative name, website, and source of identification. The process described below is what we have now, but in the beginning it was less well organized because we did not have the Gate and relied a lot on the dashboard and manual work, with messaging, telephone, and videoconference exchanges.

The Gate scrapes the public web for a first description, and an LLM condenses it into a short abstract, enough for the team to judge whether the initiative is worth studying. The team validates or not, and the Gate opens an initiative file and logs the status change (“Proposed” to “Validated” or “Non-validated”). Initiatives not selected, or that decline, are kept in the Corpus, not deleted, so that conversion rates can be calculated. The sampling frame is itself data: keeping the rejected and the unreachable is what allows reasoning later about coverage and bias.

## 5.3 Enrolling and preparing

Once an initiative is validated, the team assembles the people and permissions needed to visit it. It looks for a suitable interviewer nearby, makes contact, collects availability and chooses one; no contract is signed until the case is confirmed. In parallel it contacts the initiative to gauge interest and obtain dates and an address. When both sides are willing, the case is assigned and the interviewer contracted. The contract includes the Gate-generated case description and objectives, so that the interviewer can prepare. The more the interviewer knows about the case, the better the interview and the richer the data: the interviewee adapts to the level of the interviewer. An ignorant interviewer gets superficial answers; an informed one asks better questions, and the interviewee, more interested, volunteers expert detail.

A micro-level case is normally done by a local interviewer engaged and trained by the central team. Meso-level initiatives require more institutional legitimacy on the interviewer's side, and thorough preparation. The CEO of a multinational or a senior civil servant cannot be interviewed by a local interviewer; the PIs do it themselves, or send a seasoned interviewer with a personal introduction.

The Socioscope central team (not the interviewer) coordinates and confirms the appointment. Two consent forms were generally offered - one for research use, one for communication and the platform, with the second one applying only when participants opted in to the communication and platform components. Equipment (e.g. video camera) is dispatched and readiness checked. The case moves through the stages of Interviewer Assigned, Interview Confirmed and Interview Guide Generated. The central team confirms the appointment so that it, not the interviewer, is identified as the principal for later exchanges (e.g. video clearance). Otherwise the initiative knows only the interviewer, who may have left the Socioscope by the time of analysis. We had not anticipated this at the start, and it made it difficult to reconnect with some initiatives later.

## 5.4 Interviews: initial contact, interview and debrief

The interviewer travels to the site and conducts a long, open interview recorded in audio and video, inviting the interviewee to speak freely in their own terms. They also record a short video for the social contract, film the setting, and gather supplementary material: documents provided by the initiative and a scrape of its website. The short video follows a standard series of questions, to make editing easier, and is offered to the interviewee as a gift. All the collected material is uploaded to the Gate, which authorises the upload, provenance-stamps, renames the files, and writes an audit log (status Interview Materials Uploaded). During the proof of concept we recorded everything on video. Now only the facility tour and the short interview are filmed; the long interview is audio only. This spares interviewers the upload of massive video files, a problem where internet is slow, and spares us processing files that, in hindsight, added little.

After upload, the Gate transcribes the material and compares it against the topic guide and the case-specific questions prepared for this case. From that analysis it generates a targeted debrief interview guide: a list of what is missing, unclear, or unverified. The debrief then takes the form of a human (or an LLM) interviewing the interviewer based on that analysis - working through the gaps rather than collecting free-form commentary. Where a gap can only be answered by the initiative itself, the interviewer is asked to return to the case, by call or revisit, with very specific questions - as was done in the Postobón case. Because this return is sometimes necessary, its possible occurrence is written into the interviewer's contract: interviewers agree from the outset that incomplete data entails contacting the case again. This has two virtues - it makes interviewers respect the topic guide more closely, and it lets the initiative be warned, at the time of the visit, that a short follow-up may be needed.

After the visit, the central team in Paris held a debriefing session with the interviewer. Debriefing is an extension of interviewing, somewhat like the supervision routinely practised in sensitive relations between professionals and their patients or clients. Besides checking that the information was fully gathered, it lets someone external to the interview assess the data and act as quality controller. As the same debriefer usually supervised several interviewers, the debriefs also yielded contextual information.

Transcription and quality control of the audio must be fast, otherwise the debrief is delayed. This happened at the start, when the IT team was solving technical issues on the fly and the pipeline clogged. We advise teams attempting comparable projects to sort out that part of the chain before scaling up collection.

### 5.5 Upload, formatting and processing

As soon as interviewers collect data, they log into the Gate and upload them into the case folder assigned to them. Uploaded material first passes a basic file check (correct format, plausible size), then is renamed according to the file-name convention:

COUNTRY ID -- CASE ID – DATE – FILE TYPE – FILE FORMAT – FILE NUMBER--VERSION ID-- CONTINENT ID – LANGUAGE ID – EXTENSION

An example file name based on this convention is: FR-002-2025-12-17-SUPPLEMENTARY_MATERIAL-CSV-1-0.0.3-EU-FR.mp4.

Depending on the file type, we then proceed to transcription and translation, with transliteration where necessary. Named entities in transcripts need special handling: correction for consistency, protection from accidental translation, and checks against the initiative’s website (the vocabulary built when the website was scraped at scouting is injected to help transcribe proper names and acronyms correctly). At this stage the Gate writes provenance details into each file, such as the models used, and handles specific requests (anonymisation, metadata extraction, quality-control reports) so that the same files are not re-run repeatedly.

Less common languages may be poorly transcribed by the major engines and require shopping for the right one. In many areas people mix languages (English and Swahili, French and dialectal Arabic). Processing therefore varies by language: for languages with small training sets, such as Afrikaans, we used different transcription and translation tools than for French or English. Speech-to-text engines now cover most, though not all, of the languages we met (Radford et al., 2023).

File naming must be thought through: with hundreds of files, approximation is a recipe for problems. The files from this step are added to the Corpus, and relevant snippets (from interviews, for instance) can be used to update the initiative's page on the public website.

In hindsight, it is essential to be exacting in collecting and storing all metadata at this stage (who downloaded which files when, whether the consent forms are fully readable, and so on). Each mistake, missing value or oversight triggers a long inquiry later to fill the gap.

### 5.6 Quality control and payment clearance

Quality control is not a final gate but runs the length of the pipeline, because errors cascade if they are not caught early. A misspelled proper name, for instance, appears under several spellings at anonymisation, which then tags several entities where there is one.

Control begins with the selection and validation of initiatives, continues through the site visit and the completeness of uploads, is reinforced by the recorded debrief, and extends into processing through metadata management and checks on transcription and translation. These checks go beyond the surface to consistency, for instance in the attribution of speech turns (an interviewer who appears to speak as an interviewee).

For the first batch of 300 cases, the project ran on the order of fifty control points per case, overseen by dedicated quality-control staff and supported by regional hubs, with an independent quality check and commentary on the completeness of each case. Gradually more of the quality control was transferred to LLMs, and we consider now transferring even more parts of that process to LLM. The central dashboard, and now the Gate, tie this together, flagging cases that depart from the protocol for audit. We are still improving the process as analysis reveals new kinds of issue, and other teams will meet issues specific to their own material. This is why we strongly recommend a proof of concept to test and polish the pipeline before scaling up: a minor problem takes serious dimensions at scale. The field incidents described under Lessons learned, such as missing consent forms, illustrate what these controls catch.

Quality control gates the payment of interviewers, video editors and others. An invoice does not release payment by itself. It triggers an AI-generated quality-control report on the interview, produced from the debrief and the uploaded material, which goes to the PIs (or their delegate) for clearance. Payment follows confirmed quality, not completion of upload. Interviewers can see the quality assessment of their uploads in the Gate, so they know when to expect payment; formal scoring rules and cut-offs are still being refined. Receipt of the debrief material is acknowledged to the interviewer, with what happens next and when. International payments can be slow, and an acknowledgement of reception helps.

## 5.7 Fulfilling the social contract

Alongside the curation chain runs the return promised to the initiative and other contributors: payment of the interviewer, a short video, and a web page for the initiative. The video strand caused a disproportionate share of problems in Phase 1, so its steps are set out here in more detail.

First, hand-off: the interviewer's short-video footage and photographs are passed to the video team with a brief, e.g. which parts of the rushes might provide good illustration of the facility or site. The video team assigns videos to editors based not only on availability, but also on source language fluency. Second, editing: the editors assemble and cut the material. Editors report progress at defined stages, like rough assembly, first cut, final cut, giving each video a status that mirrors the case dashboard. Third, an automated AI quality-control pass checks the mechanically checkable: framing, audio levels, captions and subtitles, length, branding seal, obvious defects. Fourth, a human review at team level, before anything leaves the team, to avoid sending something the initiative might not like. Only then, fifth, is a draft shared with the initiative for feedback, handled by the Socioscope team, not by the editor or interviewer. Feedback goes into a final cut, published for example on YouTube and linked to the initiative’s page on the Socioscope website.

How much of this can be automated? In 2026, much of the mechanical work, but not the judgement. Current tools reliably handle captioning and subtitling in dozens of languages, transcript-based editing, silence removal, reframing, B-roll suggestion and diarisation (splitting audio into speaker-labelled segments). What still needs human hands is the creative assembly (the story, the emphasis, the tone) and a review before anything reaches the initiative. The realistic design is AI-assisted editing with human conception and review: the tools do the repetitive work and the checking; humans judge the cut that represents the initiative to the world.

The web page follows a similar discipline. The initiative’s page is drafted around a short AI-generated abstract of the collected material, edited to take away all details that can appear privacy sensitive (for example, that the interviewee sold their house to get the starting capital, or that they were illegal when they started). The page passes human validation at Socioscope team level before being published (status Web Page Created), and is sent to the initiative for review and edits. Thereafter the initiative owns its page and can edit it.

A thank-you and a small goodies package close the visit, and a final message invites the initiative to keep participating, with the sending of a participation certificate they can frame and display. The web page includes a “contact” button where other cases can send a message to the case, and the case has the possibility to display what they offer, e.g. products or services, and what they look for. These are links to their own website: no publicity or commerce is allowed on the Socioscope website. This strand is the concrete form of the social contract described above: visibility, community and recognition in exchange for deep access.

Note that the Socioscope’s social contract is project-specific. Other teams, and we ourselves later, may have to design another one. The video clip as a gift worked well in 2024-2026; it may not be enough in the coming years, when making videos becomes a widespread skill.

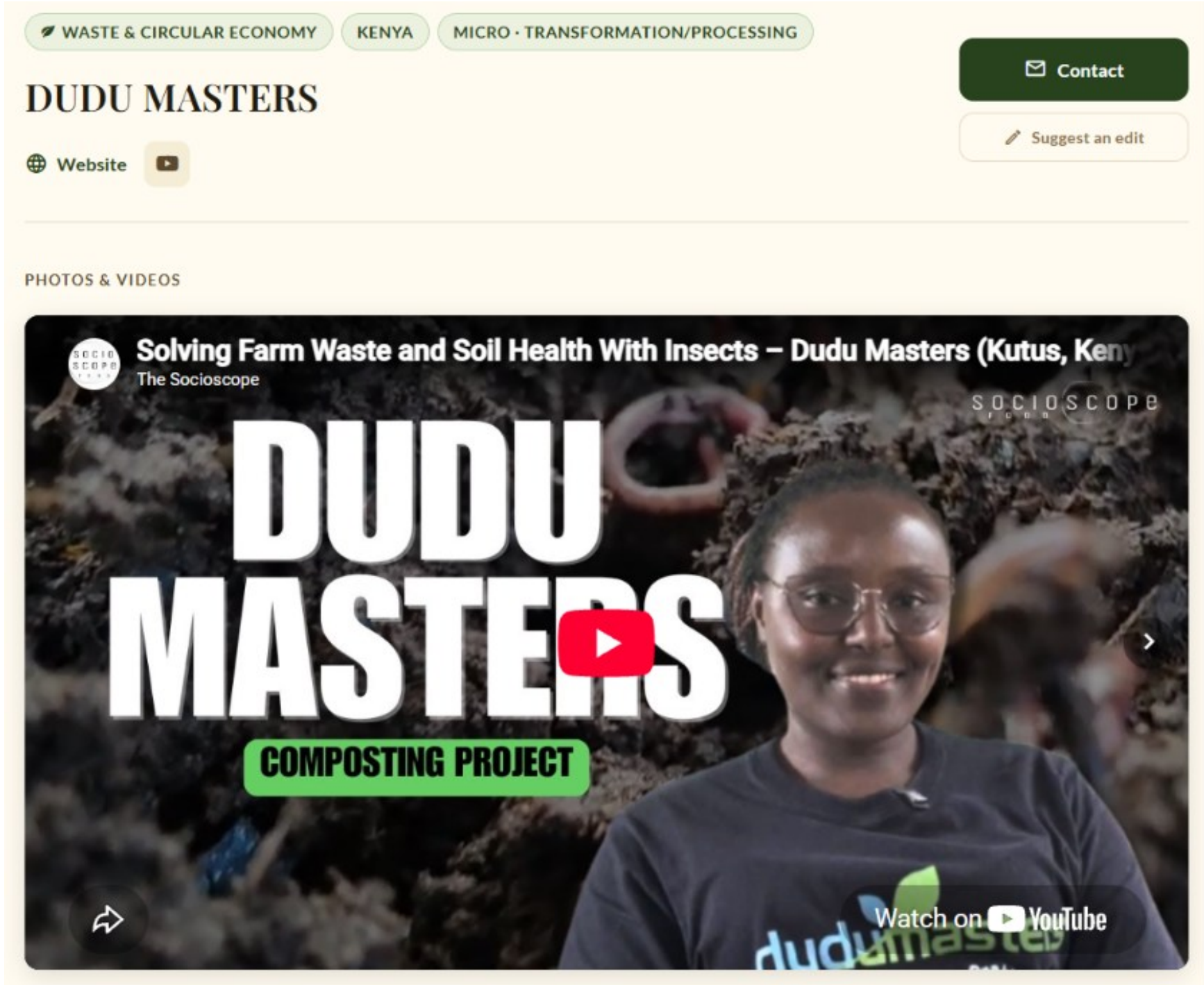

ABOUT

Dudu Masters is an innovative initiative focused on sustainable agriculture through the rearing of insects, particularly black soldier flies and red worms. Founded by Jennifer, an entomologist by training, the project transforms organic waste into valuable resources such as organic fertilizer and animal feed. Dudu Masters also offers training and starter kits to empower farmers and households, promoting a circular economy that improves soil health and food security while reducing reliance on synthetic fertilizers.

*Figure 6. Example of an initiative page on the Socioscope website.*

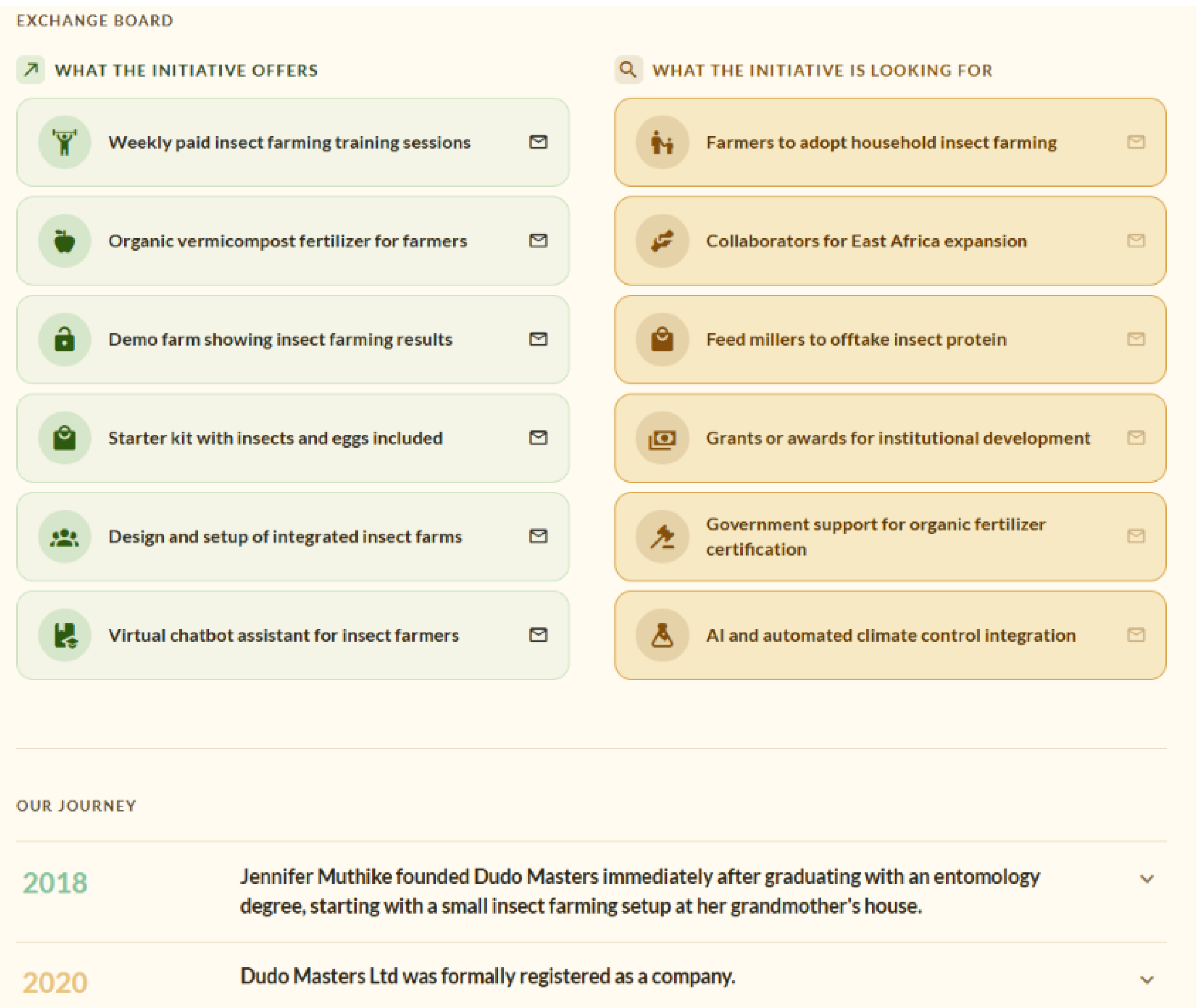

*Figure 6. Example of an initiative page on the Socioscope website (continued)*

# 6 Variations and costs of the pipeline

What is described above is our recommended configuration: a hybrid pipeline in which humans and AI each take the steps they do better. The pipeline can be run in three configurations (see Appendix 2), and the trade-offs deserve to be made explicit, because the choice is not simply “AI is cheaper”:

1. A **human-only** version, in which every step is done manually
2. A **tech-maximal** version, in which every step that *can* be automated, is
3. A **hybrid** version which we run in practice, in which each task is assigned to whichever approach the task itself favours.

One caveat applies to every figure that follows. What we cost here are marginal costs of data collection: what one more case adds, once the installation exists. They are the right numbers for deciding how to run a pipeline, and the wrong ones for budgeting a project. The fixed and shared costs are taken up at the end of this section.

At the usual operating scale (100–500 cases), and with current tools, the hybrid costs about a third less than the human-only version, while keeping human the steps where automation would cost data quality, consent quality or the relationship with the initiative.

We would not run the tech-maximal version. The per-case savings from AI come almost entirely from administrative or mechanical steps (transcription, translation, file naming, tracker updates, scheduling), where AI produces the same or better output for a fraction of the cost, with no loss that quality control can detect. Automating the transcription and diarisation of interview and debrief, for example, cuts the cost of those two tasks by about 96%; the hybrid takes that saving in full, since nothing in the interview's content or consent depends on who transcribes it.

The same costing shows the limit. First contact with a candidate initiative (a coordinator makes personal contact, explains the project and secures the initiative's interest) costs €28 of coordinator time in the human version and formally €0 in the tech-maximal one. We do not take that saving, because misunderstood information costs far more than €28 downstream. More generally, where the output depends on judgement, rapport or accountability that AI cannot yet provide (first contact, the interviews, sampling and go/no-go decisions, final quality sign-off before payment), we recommend humans.

The dominant marginal cost of deep data is human fieldwork, which is also what makes the quality of the data. Interviewers work under service contracts paying an all-inclusive fee per completed case, covering, with few exceptions, scouting, travel and field expenses. The fee is set by a budgeting instrument: a baseline of €20 per hour, adjusted by a regional multiplier (from 0.55 in South America and Africa to 1.65 in North America) derived from local living costs and local interviewer labour markets, applied to the four and a half person-days budgeted per case, plus a 25% overhead. Per-case fees range from about €400–600 in South Asia, Africa and Latin America, to €1,000–1,300 in East Asia and North America, €1,500 in Western Europe and €2,000 in Switzerland. Payment is released in tranches (typically 70/30) at the validation points described in Quality control and payment clearance, and reconciled case by case in per-country payment trackers. Local subcontractors hold the local contracts, keeping the scheme compatible with labour law in each geography. In effort, four and a half person-days per case put the 686 Corpus cases of Phase 1 at about 3,100 contracted field person-days, some fourteen person-years, before central coordination (four full-time roles over two collection years, about 2,000 person-days) and video editing. Appendix 4 details the estimate and its assumptions. These figures come after a steep learning curve during the proof of concept, which yielded 60 cases not included in the 686.

These rates are the third iteration of a costing that began with the proof of concept, where a case cost about €1,000 per interview (€800 per interview plus €1,600 for a synthesis note on a package of eight) and each recorded debrief about €200. Downstream of fieldwork, costs drop sharply. Machine transcription of the Corpus's 1,430 hours is cheap and fast (the 137.75 hours of recordings in the project's documentation library were transcribed in one working session); the remaining human costs are quality-control review and video editing (€400–600 per case in the proof of concept). The economics of deep data are front-loaded: the money goes into people travelling to places and talking to people. That is why the pipeline is built to make this expenditure permanent: data paid for once must remain analysable as tools change. Of the four and a half person-days budgeted per case on the interviewer's side, the recorded material itself (about an hour and a quarter of interview and just under an hour of debrief) is barely two hours; the rest is scouting, negotiation, planning, travel, filming, data management, upload and exchanges with the central team.

The costs above are marginal costs of data collection, not the cost of the project. They do not take into account a series of specific cases where we had to solve issues and repair errors or missing data, which is typical of a frontier research discovering unexpected problems; nor the fixed and shared costs that make the marginal cost possible: PI and co-PI time, the technical team that built and runs the Gate, the SDK and the AI processing chain, IT and storage, quality control and supervision, legal, financial and administrative support, recruitment and training, and the design of the instruments themselves. Etc. These should not be underestimated. Appendix 4 gives the order of magnitude: some 3,100 contracted field person-days for 686 cases, against 5,800-6,000 person-days for the data-collection operation as a whole, and that larger figure still excludes the technical and AI team, the PIs and institutional support. Fieldwork cost per se is therefore barely half of the effort counted in person-days, and less than half of it in money, because the rest of the installation is staffed at Paris rates. Anyone budgeting an LSQR project should read the per-case fee as one half of what a case really costs.

# 7 The Socioscope pipeline implementation: primary data and documentation

This section describes what the pipeline produced in two and a half years: the data collected (the Corpus) and the documentation needed for maintenance, provenance, quality control and exploitation, from consent forms to technical documentation and administrative records.

## 7.1 The Corpus

The pipeline is best read as a funnel. Nearly 4,000 initiatives were scouted; 3,247 entered the master log as leads; 45% of leads passed validation, and about 50% of validated cases reached completion, yielding the documented set of over 600 (686 in Corpus 0.1.0; see below).

Conversion improves as a territory matures, at the risk of over-representation: it reached 67% in Colombia. Monthly closures peaked at 110 in June 2025, a surge that outran the processing capacity of the five-person central data-collection team and forced a deliberate slow-down of field operations in the closing months, painful for field management, to keep quality control and the remaining budget on track. Once a case enters the log it moves fast: the median time from entry to closure is two months, and 70% of dated cases close within three (Appendix 4). Attrition is not waste but selection: most of the gap between scouted and completed is the central team rejecting cases that do not meet the criteria, or of which we already had enough (“enough organic farms”) -or because we realized we were going to overshoot our target.

The quality metrics of the released Corpus are Gate-based. Every recording travels with quality flags raised by the controls described in Quality control and payment clearance: unreadable or corrupted segments, speaker-attribution inconsistencies, divergent spellings of named entities. A case cannot be released while flags remain open.

Contrary to our design, we very rarely sent interviewers back to the field for missing data, simply because the pipeline was clogged. This was a serious blow to our intentions. We advise teams attempting comparable projects to watch the conversion rate closely: the learning curve is steep, and once everything falls into place the yield can overwhelm a chain designed for a slower pace.

We therefore ran a systematic, independent review of each case on some 39 points (a completeness instrument, distinct from the fifty in-pipeline control points described under Quality control). The project’s quality controller re-read every completed case package, scored each point as covered, partly covered or missing, sent comments back to the fieldwork managers, and flagged cases for re-collection. The results are summarised under Limitations.

### Data Collected

As of July 2026, the latest analysable Corpus (version 0.1.0) covers 686 cases in 31 countries. Collection concentrated on ten primary countries - Colombia (162 cases), France (80), Argentina (59), Kenya (50), Denmark (45), Peru (41), Spain (36), Costa Rica (34), Mexico (30) and Ecuador (28) - with a further 121 cases spread across 21 other countries, from Armenia to New Zealand. Some 40 cases are in the incoming pipeline waiting for completion (e.g., missing consent form, video under revision…).

| Country | Cases |
|---|---|
| Colombia | 162 |
| France | 80 |
| Argentina | 59 |
| Kenya | 50 |
| Denmark | 45 |
| Peru | 41 |
| Spain | 36 |
| Costa Rica | 34 |
| Mexico | 30 |
| Ecuador | 28 |
| 21 other countries | 121 |
| Total | 686 |

*Table 2. Distribution of closed cases by country in the Corpus, July 2026.*

The language distribution reflects the sampling: Spanish accounts for about half of the speech turns (51%), English for a third (32%), French for 12% and Danish for 3%, with a dozen further languages - Swahili, Russian, Polish, Dutch, Georgian, Cantonese, Japanese, Tagalog among them - making up the remainder.

**What is inside each case. The material is substantial:** a filmed setting and facility tour, a long audio interview, a short video interview based on identical questions, supplementary material scraped from the initiative's website and supplied by the interviewees, and a recorded debrief with the interviewer. And then the data documenting provenance of the raw empirical material and its subsequent transformations. The amount of raw data collected (in this snapshot of July 2026) is about 140 hours of video and some 1,290 hours of audio, plus supplementary material. The transcripts currently amount to about 450,000 speech turns and 12.6 million words in the original languages (12.1 million words in the machine-translated English layer). The debrief, an innovation of this pipeline, may in future be done partly by LLMs; we will test this in the future.

## 7.2 Internal documentation

In July 2026, the project’s working documents (95 GB accumulated over three years on shared drives) were consolidated into a curated documentation library of 1,432 unique documents. Alongside the documents, 93 audio and video recordings of team meetings, training sessions and Scientific Advisory Board sessions (137.75 hours) were machine-transcribed with speaker diarisation and filed in the same fourteen categories, which follow the life of the project from governance to publication. The project’s oral memory is thus searchable alongside its written one.

The library is distributed as follows:

| Category | Documents |
| --- | --- |
| 01 Project overview & governance | 59 |
| 02 Roles & team | 11 |
| 03 Administration | 678 |
| 04 Contracts & agreements | 81 |
| 05 Selection of initiatives | 3 |
| 06 Data collection | 36 |
| 07 Consent, GDPR & legal | 12 |
| 08 Training & field support | 324 (+31 transcribed meeting recordings) |
| 09 Processing & transcription | 15 |
| 10 Quality control | 10 |
| 11 Corpus curation & provenance | 1 |
| 12 Analysis & reporting | 24 |
| 13 Publications & presentations | 29 |
| 14 Meetings & project management | 149 (+62 transcribed meeting recordings) |
| Total | 1,432 documents + 93 transcribed recordings |

*Table 3. The Socioscope documentation library, July 2026.*

The distribution shows what running large-scale qualitative research entails. Nearly half of the library is administration proper (678 documents consisting of invoices, budgets and payment records kept at item level), and training and field support account for another quarter (324). The operational weight of the infrastructure sits in managing people and money.

The scientific tools, by contrast, are compact: 36 data-collection documents and 12 consent and legal templates, because a small set of standardised protocols serves hundreds of cases.

These counts hide very unequal objects. A consent template is a page or two; the master case-log dashboard, one spreadsheet in the count, holds 54 country sheets and 3,247 dated lead rows, one per initiative logged since 2023, each carrying the case's details and status at each step, actors and payment validation. Counting documents therefore understates the administrative mass: the paper trail is thin where research is standardised and thick where the operation deals with people and money.

The audit logic of the Corpus (see Preserving provenance) thus extends to the organisation that produced it: a reviewer can trace not only what was done to the data, but why the procedure existed, who decided it, and when it changed. Every file is registered in a checksummed manifest; duplicates are cross-referenced; exclusions are listed with reasons.

Even so, some uncertainties remain on provenance and data loss, and we still find missing or uncertain dates, hardly readable names on consent forms, a few duplicates in the supplementary material, speech turns wrongly attributed, though these get rarer as quality control cleans the data. The procedure was developed gradually, was less efficient at the start, and the documentation is not perfect.

# 8 Lessons learned and limitations

## 8.1 Lessons learned

The data collection phase succeeded, but it raised operational, methodological, institutional and logistical problems. Collecting 600+ cases across 37 countries in less than two years required continuous adaptation, close coordination and fast problem-solving.

The main problems were:

**Missing consent forms and footage**: One of the most serious issues was a completed interview without a signed consent form. In one case the interviewee could no longer be contacted; despite efforts, the team could not reconnect, and the case was excluded on ethical grounds. In another instance, an interview video was irretrievably lost when a field interviewer changed phones without backing up the original file; the interviewer revisited the initiative and conducted a second interview to recover the missing footage. The main lesson: the interviewer must have the consent form signed before the interview starts.

**Equipment and technical mishaps**: One interview was recorded with incorrect camera settings, resulting in an unusable slow-motion video; the fieldworker returned to the site and re-recorded the session under standard parameters. In another case, in Marseille, a technical failure during the interview lost the audio track. The fieldworker could not return in person, so a remote Zoom interview was organized and recorded instead, and the case could be included in the Corpus.

**Malpractice**: Despite clear instructions, interviewers did not always follow the protocol exactly. Some filmed in portrait instead of landscape, which complicated editing. Others asked interviewees to rank their stakeholders, which left the transaction grid unevenly filled. Some interviewers were simply more thorough than others. The lesson: remind interviewers continuously that the protocol must be followed strictly, and have a quality-control procedure to check that it is.

**Pace**: The initial goal was 600 case studies over two years, assuming steady growth. The process started very slow. As more interviewers joined, the process accelerated unexpectedly: the more interviewers were involved, the more motivated they were to find and document new initiatives, not least for their own financial interest. Once this snowball was rolling, the processing capacity of the central team was overwhelmed. The imbalance became acute in May and June 2025, when 101 and 110 cases were closed respectively, nearly one third of the overall target in two months. Transcriptions and debriefs fell behind schedule, which degraded the quality of the debrief sessions, and budgetary limits risked being reached prematurely. To safeguard resources for the full set of 600 cases, operations had to be slowed down. This “landing phase” required careful management of fieldworkers’ expectations, because many validated initiatives were then barred from becoming cases by the central team. This frustrated interviewers who expected to process them and had in some cases already made contact. For this reason we did not halt operations abruptly when the target was reached, and overshot the Phase 1 target of 600 cases by more than 10%. The lesson: anticipate not only delay but over-acceleration, and prepare to balance enthusiasm in the field against funding and staffing. The pipeline is long, so the effects of the learning curve, here the rise in productivity, take a while to show.

**Transcription and translation**: Most interviews were in languages readily compatible with automated transcription, but some were not. An interview in isiXhosa could not be automatically transcribed at all, for lack of speech-to-text support, and had to be transcribed manually. Other interviews in languages like Afrikaans and Danish did not get transcribed well, which resulted in downstream translation problems. We first relied on AssemblyAI and DeepL, then moved to a language- and region-specific pipeline with several providers, including Deepgram, ElevenLabs and Opus. Despite fast progress, transcription and translation quality is uneven across languages; hybrid workflows combining machine processing and human linguistic expertise are needed for accuracy and cultural sensitivity. Teams handling non-dominant languages should plan for this, including partnerships with local translators or open-source language models. Transcription must also track speech turns correctly, above all to separate the interviewer’s turns from the interviewee’s. These issues were serious at the start (2023-2024); with the progress of LLMs they have become minor.

Together, these incidents show the need for redundancy in field protocols (immediate backups, systematic quality checks, ways to re-engage a case) to protect data integrity in multi-site qualitative research. They were few relative to the number of cases, but they show why quality assurance and close communication between the central team and interviewers matter. The design implications of these lessons are consolidated in Appendix 1 (Before attempting LSQR: sixteen rules in five families).

### 8.2 Limitations

The pipeline’s quality claims have limits, which should be stated. The first is that quality control is Gate-based: the released Corpus carries no unresolved flags, but we cannot yet publish rate-based metrics, such as how many errors each stage produced and caught, inter-rater agreement between quality controllers, or word-error rates of the transcription engines against human reference transcripts. Our procedures were built progressively and were not applied the same way to the whole Corpus. We made homogenising passes in May, June and July 2026, with many errors or missing data having already been corrected beforehand.

There are three partial exceptions. The first is a validation protocol for these metrics (a three-level human-versus-LLM benchmark), designed and piloted in 2024. The instrument holds 116 questions across eight cases (41 factual, 38 light-inference, 37 complex-inference), 91 with verified answer keys and programmatic scoring, and answer documents survive for three raters. Its results were never consolidated into publishable figures: the students who ran the study did not finish it. The second is an inter-rater reliability study, under way at the time of writing: several raters independently code each case on five dimensions (trigger type, geographic scope, urgency, and two presence checks), with consensus and minority ratings preserved case by case (170 cases coded as of July 2026); its agreement statistics will be reported separately.

The third: the project can already report coverage, as distinct from error. The independent 39-point completeness review described in Section 7 found that the 197 cases examined covered on average 60% of the checklist (median 61%) and flagged 13 for re-collection. A 45-case country audit (Denmark) and a corpus-wide file-level quality report (December 2025) complete the picture. So far, external assessment of the pipeline rests on its architecture (immutable originals, logged transformations, human gates) rather than on published performance figures.

The second limitation is that the analysable Corpus includes an English machine-translated layer of 12.1 million words that has not been systematically validated against human translation. Analyses run on that layer inherit whatever biases the translation engines carry. The mitigation is structural: the originals are immutable, so the layer can be regenerated and re-checked as engines improve. It is a mitigation, not a solution. We know where the engines fail: one interview, in isiXhosa, had no speech-to-text support and was transcribed by hand; machine translations from Swahili, Russian, Danish, Polish and Arabic dialects misrendered context-specific vocabulary and idiom and had to be corrected by native speakers. Some loss in translation will remain however good the instruments, since each culture represents the world in its own terms (Whorf, 1956). On the positive side, more of the expected cultural barriers were overcome by local flexibility and ingenuity than we feared.

The third limitation is structural: standardisation is a monoculture. A single transaction grid, protocol and pipeline applied worldwide is what makes 686 cases comparable; it also replicates any blind spot of the instrument 686 times. However open our protocol, LSQR meets, to a lesser degree, the problem of all quantitative research:

by standardising collection, it cuts out part of the phenomenon. Interactions the transaction grid does not capture are not recorded, and remain unknown.

The fourth limitation is inherent in the design: the Corpus is not a probability sample. Its geography reflects where partner organisations and interviewers could be contracted, and how each progressed. We followed the scent of promising initiatives, as defined by our criteria, and took opportunities where they arose. Colombia alone accounts for nearly a quarter of the cases, and ten countries for over four-fifths. Participation is voluntary and mediated by the social contract. We were looking for initiatives active in more sustainable practices and thus willing to be visible. This bias is partly offset by the confidential layer, and by the fact that interviewees speak at length about actors who did not, or were not chosen to, participate (see Data extraction).

Finally, the pipeline described here is a snapshot of a fast-moving chain (see Conclusion). The durable claims of this paper are its principles (provenance, gates, immutability), not the tools, which will soon be replaced.

## 9 Conclusion

This paper has described, at the level of procedures, what it takes to collect and curate qualitative data at scale: a systemic sampling frame operationalised by the transaction grid; a social contract that rewards interviewees and grants access; a pipeline that moves every case through scouting, enrolment, collection, formatting, quality control and curation, with a status and a log at each step; and a provenance discipline that keeps originals immutable and every transformation traceable, applied alike to each case, to the Corpus, and to the project's own documentation.

Concept extraction at scale, post-coding a very large corpus, is possible with off-the-shelf high-end LLMs and some programming. Some colleagues doubted, and so did we, that a machine could do the fine abductive work of a qualitative researcher. That was settled early and clearly. Using AI also changes the way we do research, with new opportunities and new problems. A coding grid or a concept can be tested across a large sample in a couple of hours, rather than the months, if not years, it would take by hand. Changing analytic lens and testing its validity becomes cheap, which accelerates recursive and abductive engagement with the data, in which surprising observations prompt new conjectures that are then checked against the corpus. We can apply more freely the abductive process that is the core of data interpretation (Lahlou, 1995, 1996), and compare several interpretation trajectories. The difference is one of quantity, but large enough to change what can be attempted. It is liberating.

This has a cost. Scaling up is not “qualitative work plus a text-mining step”. It is a different undertaking, organisationally and epistemically. It requires industrialised data collection, structured data management and quality control, and continuous technical updating. Ten months into setting up the AI processing, the team was on its tenth prototype of the chain, having used many different LLMs at different steps. The lesson is structural: the data are the stable asset of the research, the techniques keep changing. The pipeline must therefore be built to preserve and re-process the data as tools evolve, a requirement the FAIR principles articulate as findability, accessibility, interoperability and reusability (Wilkinson et al., 2016).

This approach yielded 686 cases from 31 countries, some 1,430 hours of recordings and 12.6 million words of transcript in under three years. We also hold video material of similar magnitude, whose analytical use beyond illustration remains to be explored. The result is, to our knowledge, among the largest systematically collected qualitative corpora.

The analysis under way is building typologies and ontologies of actors and transactions, testing on hundreds of cases hypotheses formed on a few, and relating the micro, meso and macro levels. Variables not envisaged at the start are still being found, and the corpus will answer questions that were not asked when it was collected. That is what a preformatted instrument cannot give.

Three lessons stand out. First, scale in qualitative research is an organisational achievement before it is a technical one: the binding constraints are not storage or compute; they are to access and keep good data sources, to train and coordinate people, to secure consent, quality control and payment at every step. Second, the pipeline is what makes the corpus scientific rather than merely large. Every file carries its provenance; any analysis can be traced back to an immutable original, re-run and audited. That is the condition for cumulative and reproducible qualitative research, and the condition on which social science can take up what AI offers at scale without giving up its evidentiary standards. Third, in an age of synthetic and fabricated data, where material is cheap to generate and hard to authenticate, the provenance, depth and curation of real data are what separate evidence from plausible text. LSQR must be built on that ground. It is costly. It is also feasible, and on three years of evidence, worth doing.

## Glossary

[some entries are copied or adapted from (Lahlou, 2017, 2024).

*affordance (of an object for a subject):* This object's potential for action (for this subject).

*case:* The data collected about a data source (an initiative), digitized, documented and archived in the Corpus.

*Corpus:* The digital repository of all the cases. It can only be incremented: originals are immutable and nothing is erased.

*data pipeline:* The data journey from the moment a source is scouted to the moment the collected data sit, in the right format with adequate labelling and metadata, in the Corpus.

*data provenance:* The documented history of a data item: where, when and how it was collected, and every transformation applied to it since.

*debrief:* A recorded interview of the interviewer, conducted after upload, on the basis of an analysis of the collected material against the topic guide. It works through what is missing, unclear, or unverified.

*embodied competence:* The ability by which the perception of a given object or situation triggers a relevant action.

*Gate (the):* Our proprietary tool that stores the Corpus, restricts and logs access to it, and documents every operation made on it.

*initiative:* A real-world entity (programme, organisation, device) that tries to move the system in a given direction. The initiative is the sampling unit; the case is the material collected about it.

*installation:* Specific, local, societal settings where humans are expected to behave in a predictable way. Installations consist of a set of components that simultaneously support and control behaviour, distributed over the material environment (affordances), the subject (embodied competences) and the social space (institutions).

*Large-Scale Qualitative Research (LSQR): Collecting open-ended qualitative data on many cases with a single protocol, and using AI to make the material analysable at scale.*

*panelisation:* Selecting cases for repeated data collection over time. A panelised case can be asked new questions as the research questions evolve.

*role (of a person):* The set of behaviours that others can legitimately expect from a person.

*SID (Socioscope ID):* The unique identifier assigned to every speech turn in the Corpus. It allows any finding to be traced back to its verbatim.

*snapshot:* A frozen, immutable, citable state of the Corpus, on which a given analysis rests.

*social contract:* The combination of a role and a status ('you do this, you get that'). In the Socioscope, the balanced exchange with participants: deep access in return for the video, the web page, community membership and recognition.

*status (of a person):* The set of behaviours a person can legitimately expect from others.

*system:* A set of interrelated entities whose interrelationships enable the system to be maintained as it is, with some resilience.

*transaction grid:* A data-collection tool that lists every stakeholder an initiative transacts with, and what each party gives and gets in each transaction, from both perspectives when possible.

*zone of interest:* A territory where sampling was deliberately densified, to get a more complete view of the local system.

## Acknowledgements

**This research project and related results were made possible with the support of the NOMIS Foundation.**

We thank the fieldwork managers - Sofía de Vega, Julia Movshovich and María Mejía - who ran the pipeline's daily operations; Antoine Cordelois (CTO) and Do Huynh (AI engineering), who built and operated the processing and data infrastructure; Johannes Oster for his contributions to the tech infrastructure; Alex Cano, the project's dedicated quality controller; Emmanuelle Honoré (fieldwork manager) and Loïc Bonin (assistant fieldwork manager), who ran the proof-of-concept phase; the regional coordinators - Romina Sarmiento, Leonardo Herrera, Cathy Kamanu, Ravenn Tripplet, Abril Campos, and Keshia Hoaeane - and the partner organisations behind them; the more than eighty interviewers who collected the cases; Hugo Saugier, Karen Garcia, Ricardo Talens, Celine Gurteen, Juliette Capus and the video-editing team; Barbara Grassauer, Nadège Bourgeois and Irina Nosulenko for administrative support; Mathieu Jacomy and Anders Kristian Munch for their advice on AI and data analysis; Atrina Oraee, Jinyu Cong and Jeevya Aroun, who ran the 2024 LLM validation study; the Scientific Advisory Board; and the Ethics and Deontology Board of the Paris IAS. The project-side personnel are listed in Appendix 6.

## Author contributions (CRediT)

Roles follow the CRediT taxonomy, in its own order; the parentheses give the specifics.

Saadi Lahlou: conceptualization; formal analysis; funding acquisition; investigation (fieldwork); methodology; project administration; supervision; writing - original draft; writing - review and editing. Principal investigator.

Juan Pablo Caicedo: data curation; investigation (fieldwork); project administration (data collection; oversight of the fieldwork managers, of video production and of communication; legal and contractual arrangements); resources (design and management of the project documentation and procedures); writing - review and editing.

Shriya Sekhsaria: data curation (version ledger; project historization); methodology (AI processing chain); software (design and coding of the Gate and of the internal and external websites); validation (2024 human-versus-LLM benchmark; inter-rater study); writing - review and editing.

Valentine Fournand: data curation (design and maintenance of the case log and of the central dashboard); investigation (fieldwork; synthesis of field feedback); project administration (oversight of the fieldwork managers, of the interviewers, of video production, of communication and of the websites); resources; writing - review and editing.

Paulius Yamin: conceptualization; investigation (fieldwork); methodology (field protocols and topic guide); project administration (budgets and management); resources (interviewer training and field-operations material); writing - review and editing. Co-investigator.

Helga Nowotny: conceptualization; formal analysis; funding acquisition (funding and partnerships); investigation (fieldwork); methodology (research design); supervision; writing - review and editing. Principal investigator.

**Declaration of generative AI use in manuscript preparation**

Beyond the AI processing that is itself the object of this paper (transcription, translation and extraction within the pipeline, documented in the text), the authors used Claude (Anthropic) as an editorial assistant in preparing the manuscript: verifying reported figures against the project's master case log and documentation library, checking internal consistency of numbers, cross-references and terminology, suggesting copy-edits and candidate references, and drafting some figures. All AI-suggested changes were introduced as tracked revisions and individually reviewed, accepted, rejected or rewritten by the authors; all suggested references were verified against the original sources. The research design, analyses, interpretations and final wording are the authors' own, and the authors take full responsibility for the content of the publication.

## Appendices

The Data Collection Protocol (V8, 55 pages) is available from the authors on reasonable request, together with the four training videos that go with it. Its table of contents is reproduced as Appendix 7. Appendices 1–7 follow.

## Appendix 1. Before attempting LSQR: sixteen rules in five families

Each rule below is developed in the body of the paper. The rules should ideally be settled before attempting a comparable project; the table at the end of this appendix summarises, for each rule, when it must be decided, what it costs to get it wrong, and our evidence.

**●1. Size and shape are design parameters, not dials**

Scale is not something a project acquires gradually. It is chosen at the outset, and the choice determines what kind of organisation is required to deliver it. Everything downstream - budget, team, technology, contracts - is a consequence of this first decision.

**●1.1. Determine the size of the endeavour before costing it**

More cases are cheaper per case and more expensive in total: unit cost and total project cost move in opposite directions, and both must be on the table before starting. Size also changes the management regime discontinuously. Fifty cases in a single country and a hundred or more across several countries and languages are both LSQR. A sizing decision taken implicitly is a management model adopted by accident.

**●1.2. Derive the technology/human split from size and research objective**

The share of the pipeline that is automated is not a matter of taste, nor of which tools happen to be available. It follows from how many cases will be collected and what the research must ultimately extract from them. The split decided here propagates down the whole pipeline: it sets transcription capacity, shapes quality control, determines the skills profile of the people recruited, and investigator involvement in management and key decisions.

**●1.3. Prove the whole chain at tens before committing to hundreds**

Test the full chain - interviewers, equipment, transcription, upload, quality control - on tens of cases before scaling to hundreds. A minor problem takes serious dimensions at scale. The proof of concept produces two things nothing else can. First, a measured unit cost and cycle time, which converts the sizing assumption into a defensible budget. Second, clarity about what the analysis will actually need: our own first months were slowed by exploring which data were required, because at scale a question that was not asked cannot be re-asked. Extraction targets must be fixed before the field opens, and the proof of concept is where they get fixed.

**●2. Contract the future, not the fieldwork**

At scale, contracts are infrastructure. Procedures can be corrected mid-flight; contracts cannot be renegotiated once the data exists. This family covers everything that has to be written down before the first piece of data is collected.

**●2.1. Negotiate data retention before collecting**

The default three-to-five-year period silently caps the useful life of the corpus; we obtained fifteen. The decision is taken once, before collection, and it fixes how long the dataset can be exploited - including by researchers who are not yet on the project.

**●2.2. Build the legal and ethical infrastructure ahead of the field, at the scale of the project**

You are working with people, and the obligations that creates do not scale by improvisation. Ethics committee approval, privacy regimes that differ by geography, named data controllers and processors, non-disclosure and confidentiality clauses for staff, interviewers and partners, and a breach procedure must all be drafted, approved and in place before collection opens - and sized for every jurisdiction in the plan, not only for the country of the coordinating institution. This is the direct counterpart of 2.1: the longer the data is meant to live, and the more territories it comes from, the more of this has to be right the first time.

**●2.3. Design the social contract, and its logistics**

Decide what participants get in return, and treat delivery - videos, web pages, certificates, shipping - as a supply chain rather than a courtesy. Expect its currency to change over the years: what was attractive to an initiative at the start of a multi-year project is rarely what is attractive at the end.

**●2.4. Write the follow-up into the contract**

Interviewers agree from the outset that incomplete data means contacting the case again, and the initiative is warned at the visit that a short follow-up may be needed. Retrofitting this obligation once fieldwork is under way is expensive and frequently impossible.

**●2.5. Gate payment on quality, not on delivery**

An interviewer's invoice triggers a quality-control report; payment follows confirmed quality, not completion of upload. This single clause does more for data quality at scale than any amount of downstream correction.

**●3. Keep the core central, and partner for the rest**

Scaling is an organisational problem before it is a logistical one. The question is not how many interviewers can be hired, but what must remain in the hands of the central team and what can be delegated to institutions or people that are already present where the data is.

**●3.1. Make the central team the principal**

The central institution, not the interviewer, confirms appointments and owns the relationship with the initiative. Interviewers leave; the relationship must survive them, because the initiative will be contacted again - for follow-up, for validation, and in later phases of the research.

**●3.2. Find partners to operate at scale**

A central team cannot itself be present in every geography, and should not try. Decide what is genuinely central - the protocol, quality control, the relationship with the initiative, and ownership of the corpus - and find the right partner for everything else. The form of the partnership follows the size and objective of the project: shared ownership of the data, granted rights of usage, or straightforward paid collection. Settle the terms before collection begins; renegotiating them once the data exists is a different and much weaker conversation.

**●3.3. Test interviewers, and train for teamwork**

Experienced academics often do not follow protocols: of thirty tested in our proof of concept, very few were retained. Selection must be empirical rather than by credential, training must target teamwork and protocol compliance rather than interviewing skill alone, and compliance must be controlled continuously rather than assumed.

**●4. Manage throughput, not speed**

Once the pipeline is running, the binding constraint is not how fast the field can work but how much the central team can absorb. These rules are tunable week by week - but only if the pipeline was instrumented from the first day.

**●4.1. Put the automation line where the configuration requires**

Prioritize automating the mechanical steps - transcription, naming, tracking, with savings up to some 96% per step - and keep humans where judgement, rapport or accountability matter: outreach, interviews, go/no-go decisions, final sign-off. This is the operational expression of the split derived in 1.2.

**●4.2. Sort transcription and translation before scaling, especially on small languages**

Small languages need engine shopping and native-speaker review, and mixed-language speech is common in the field. Transcription is the throat of the pipeline: a slow chain there clogs everything upstream of it, including the interviewers waiting for quality clearance and payment.

### ●4.3. Plan for over-acceleration, not only for delay

Field enthusiasm can outrun central processing capacity. Cap monthly case closures at what the central team can absorb; an uncontrolled surge produces a backlog that degrades quality control and delays payment, which in turn damages the field relationships that produced the surge.

## 5. The corpus is the product

The deliverable of a data collection phase is not a set of completed cases but a dataset that someone who was not there can use years later. The rules that make that possible cost almost nothing on day one and cannot be retrofitted.

### ●5.1. Fix file naming and provenance on day one

Immutable originals; every transformation logged with the model and version that produced it; explicit, predictable file names. Approximation is survivable at tens of files but is a guarantee of problems at hundreds.

### ●5.2. Track the funnel from the start

Log every lead and keep the rejected ones: the sampling frame is itself data, and it is the only evidence of what the corpus is representative of. Monitor conversion throughout - ours ran at 45% from scouted to chosen, and 50% from chosen to closed case.

| Rule | When it must be decided | What it costs to get it wrong | Our evidence |
|---|---|---|---|
| **1. Size and shape are design parameters, not dials** | | | |
| 1.1 Determine the size of the endeavour before costing it | Before starting | Unit cost and total project cost move in opposite directions; a sizing decision taken implicitly is a management model adopted by accident | 600 cases across 37 countries and several languages - a different management regime from 50 cases in one country, though both are LSQR |
| 1.2 Derive the technology/human split from size and research objective | Before starting | Propagates down the whole pipeline: transcription capacity, quality control, skills recruited, and investigator involvement in management and key decisions | We did 200 cases of human quality control which was more expensive and clogged the central team sporadically. |
| 1.3 Prove the whole chain at tens before committing to hundreds | Before the field opens | A minor problem takes serious dimensions at scale; a question that was not asked cannot be re-asked | Tens of cases tested end to end; our first months slowed by unfixed extraction targets |
| **2. Contract the future, not the fieldwork** | | | |

| Rule | When it must be decided | What it costs to get it wrong | Our evidence |
|---|---|---|---|
| 2.1 Negotiate data retention before collecting | Before the first piece of data | Silently caps the useful life of the corpus; not re-negotiable once data exists | Default 3–5 years; we obtained 15 |
| 2.2 Build the legal and ethical infrastructure ahead of the field, at project scale | Before the first piece of data | Ethics approval, per-geography privacy regimes, controllers and processors, NDAs and breach procedures cannot be improvised across jurisdictions | Ethics approval, legal bases and processor clauses covering every jurisdiction in the plan, not only the coordinating institution's |
| 2.3 Design the social contract, and its logistics | Before the field opens | Undelivered returns cost field trust; the currency changes over a multi-year project | Videos, case web pages, certificates - run as a supply chain |
| 2.4 Write the follow-up into the contract | At contracting | Incomplete cases become unrecoverable once the interviewer has been paid and released | Interviewer clause, plus a warning to the initiative at the visit |
| 2.5 Gate payment on quality, not on delivery | At contracting | Payment on upload removes the only leverage that exists over data quality | Invoice triggers a QC report; payment follows confirmed quality |
| **3. Keep the core central, and partner for the rest** | | | |
| 3.1 Make the central team the principal | At organisational design | The relationship with the initiative leaves with the interviewer | The central institution confirms appointments and owns the relationship |
| 3.2 Find partners to operate at scale | Before country launch | Terms negotiated after the data exists are a different and much weaker conversation | Shared ownership, granted usage, or paid collection - matched to size and objective |
| 3.3 Test interviewers, and train for teamwork | At recruitment, then continuously | Credentials do not predict protocol compliance; non-compliance is otherwise discovered downstream, case by case | Of 30 tested in the proof of concept, few retained |
| **4. Manage throughput, not speed** | | | |
| 4.1 Put the automation line where the configuration requires | At pipeline build; revisable | Automating judgement destroys quality; leaving the mechanical manual destroys the budget | Savings up to some 96% per automated step |
| 4.2 Sort transcription and translation before scaling, especially on small languages | Before scaling up | Transcription is the throat of the pipeline; a slow chain clogs everything upstream, including interviewers awaiting clearance and payment | Small languages need engine shopping and native-speaker review; mixed-language speech is common |
| 4.3 Plan for over-acceleration, not only for delay | Weekly, once running | A field surge produces a backlog that degrades quality control and delays payment, damaging the relationships that produced it | Monthly closures capped to what the central team can absorb |
| **5. The corpus is the product** | | | |
| 5.1 Fix file naming and provenance on day one | Day one; not retrofittable | Approximation is survivable at tens of files and a guarantee of problems at hundreds | Immutable originals; every transformation logged with model and version |
| 5.2 Track the funnel from the start | Day one; not retrofittable | Without the rejected leads there is no evidence of what the corpus is representative of | 45% scouted → chosen; 50% chosen → closed case |

## Appendix 2. The Socioscope data pipeline, step by step

This is a preliminary rendition of our pipeline, which rests on a variety of assumptions specific to the Socioscope, such as an interest in global systemic change (some researchers may have a regional focus, etc.), videos and web pages being the main social contract with participants, and a hybrid human-AI implementation.

For each of the steps in the pipeline, this table shows how the step would be done by a human, how it would be done by AI/automation, and which we actually currently use and why. The recommended approach is **highlighted** in its column. Two steps (25, 43) are the initiative's or interviewer's own action, not a pipeline design choice, and are marked accordingly rather than assigned a recommendation.

| # | Event | Human version | Tech version | Reason for choice |
|---|---|---|---|---|
| 1 | Initiative identification | **Data team** | Data team | Sourcing candidates relies on local knowledge and contacts an algorithm doesn't have. |
| 2 | Scrape web for candidate profile | Coordinator (desk research) | **Gate (scraper)** | Mechanical information-gathering; no judgement involved. |
| 3 | Draft abstract | Coordinator | **Gate (LLM)** | Condensing a scrape into a summary is low-risk and easily checked downstream. |
| 4 | Transfer abstract; assign for validation | Coordinator | **Gate** | Routing and messaging; purely mechanical. |
| 5 | Initiative validation (go/no-go) | **Research team** | Research team | Decision always human - this is the scientific judgement call that shapes the sample. |
| 6 | Register initiative file; log status change | Coordinator | **Gate** | Mechanical logging once the human decision is made. |
| 7 | Non-selected initiatives retained in corpus | **Research team / Operations** | Gate | Simple record-keeping tied to the validation decision; kept with the same team for now. |
| 8 | Search for interviewer in geography | **Research team / Operations** | Gate (algorithmic match) | Matching still benefits from local knowledge an algorithm doesn't have. |

| # | Event | Human version | Tech version | Reason for choice |
|---|---|---|---|---|
| 9 | Contact potential interviewers | Research team / Operations | N/A | Relationship-building; the workbook flags this as having no viable automated substitute. |
| 10 | Receive interviewer availability; select | Research team / Operations | Gate (confirm shortlist) | Final selection is a judgement call, not just an availability match. |
| 11 | Get interviewer's proposed dates | Data team | Gate | Low-stakes scheduling; currently handled by the data team but a reasonable future automation candidate. |
| 12 | Initial contact with initiative | Data team / Comms | Gate (templated) | First impression with the initiative; the workbook flags this as having no viable substitute. |
| 13 | Declined initiatives retained in corpus | Research team / Operations | Gate | Same as event 7 - kept with the team that made the call. |
| 14 | Notify initiative an interviewer will contact them | Internal team / ops team | Gate | Low-cost courtesy message, currently sent by the team directly. |
| 15 | Notify interviewer; formally assign case | Coordinator | Gate | Mechanical notification and status update. |
| 16 | Send offer + guidelines + case abstract | Data team | Gate (auto-generated) | A templated send; currently kept with the data team for a personal touch. |
| 17 | Set up contract with interviewer | Operations: Finance & HR | Operations: Finance & HR | Legal execution requires accountable sign-off. |
| 18 | Assign initiative to interviewer | Coordinator | Gate | System-level assignment once the match is made. |
| 19 | Give tech / standards guidelines to interviewer | Data team | Gate | Templated content; currently sent directly by the data team. |
| 20 | Send camera / tripod / mic if needed (and trace shipment | Operations | Operations | Physical logistics - someone has to actually ship the equipment. |

| # | Event | Human version | Tech version | Reason for choice |
|---|---|---|---|---|
| | for future recovery) | | | |
| 21 | Confirm interviewer readiness | Data team | Gate (digital checklist) | A quality check before the field visit benefits from a human judgement call. |
| 22 | Coordinate interview date with initiative | Interviewer | Interviewer | Relationship-based scheduling - no viable substitute per the workbook. |
| 23 | Confirm appointment to initiative | Data team | Gate | Confirmed by the central team, not the interviewer, so the initiative always has a stable point of contact (see Enrolling and preparing). |
| 24 | Share consent forms | Data team | Gate (auto-dispatch) | A sensitive legal document; currently sent with human oversight. |
| 25 | Sign consent forms and return | Initiative | Initiative | The initiative's own action, not a pipeline design choice. |
| 26 | Send reminder before interview | Coordinator | Gate | Mechanical, time-triggered message. |
| 27 | Produce specific interview guide for the case | Scientific Analyst | Gate (LLM) | Personalising a guide from case data is low-risk and reviewable. |
| 28 | Travel to initiative | Interviewer | Interviewer | Physical presence is the point - no substitute. |
| 29 | Long interview | Interviewer | Interviewer | The core relational, data-quality act of the whole pipeline - no substitute. |
| 30 | Short video interview (social contract) | Interviewer | Interviewer | Authenticity depends on a human presence on camera. |
| 31 | Collect supplementary material | Interviewer | Interviewer | Field collection tied to the visit itself. |
| 32 | Upload raw material to the Gate | Interviewer | Interviewer | The person who collected the material is best placed to confirm it uploaded correctly. |

| # | Event | Human version | Tech version | Reason for choice |
|---|---|---|---|---|
| 33 | Authorise upload; provenance-stamp, rename, log | Coordinator | Gate | Mechanical file processing once material is in hand. |
| 34 | Debrief the interviewer on gaps | Scientific Analyst leads | Gate (LLM) leads, interviewer answers | Targeted gap-filling questions don't need a human interviewer asking them. |
| 35 | Quality control of uploaded material | Coordinator | Gate | Systematic checks are more consistent run by the Gate. |
| 36 | Transcription of file | External transcription service | Gate | A 96% cost saving with no detected quality loss (Appendix 5) - one of the clearest cases for automation. |
| 37 | Generate debrief guide from gaps | Scientific Analyst | Gate (LLM) | Mechanical synthesis from a checklist of what's missing. |
| 38 | Translate transcript | Coordinator + translation agency | Gate (LLM) | Machine translation is faster and cheaper; named entities get separate human review (see Upload, Formatting, and Processing). |
| 39 | Anonymise transcript | Coordinator | Gate (LLM) | Consistent, rule-based redaction. |
| 40 | Generate metadata & audit record | Coordinator | Gate | Mechanical, systematic record generation. |
| 41 | Content QC against protocol | Scientific Analyst | Gate (LLM) | Consistent scoring against a fixed checklist; low-confidence cases still get routed to a human. |
| 42 | Commit version to corpus; log | Coordinator | Gate | Mechanical versioning and logging. |
| 43 | Interviewer submits invoice | Interviewer | Interviewer | The interviewer's own action, not a pipeline design choice. |
| 45 | Authorise payment; log | Scientific Manager | Gate | Decides a payment already cleared by quality sign-off upstream (see Quality control and payment clearance) - the judgement call happened earlier in the chain. |

| # | Event | Human version | Tech version | Reason for choice |
|---|---|---|---|---|
| 44 | Interviewer payment processed | Operations: Finance & HR | Operations: Finance & HR | Financial execution requires accountable sign-off. |
| 46 | Prepare social-contract content | Interviewer / Video | Interviewer / Video | Field-collected content tied to the visit. |
| 47 | Fulfil social contract | Operations / Comms | Gate (auto-dispatched) | Currently sent manually to keep the exchange personal; a plausible future automation candidate. |
| 48 | Send thank-you package | Operations | Gate (auto-generated) | Personal touch is part of the point of the reciprocity described in Further Enrichment with panelisation. |
| 49 | Hand material to video editor | Coordinator | Gate (auto-package) | Coordination handoff; currently manual but low-risk to automate. |
| 50 | Send first cut to initiative for feedback | Coordinator | Gate | Relationship touchpoint - the initiative sees itself represented for the first time. |
| 51 | Produce revised video from feedback | Video Editor | Gate + Video Editor (review only) | Editorial judgement; the workbook notes AI-assisted edits reduce narrative quality. |
| 52 | Upload video (YouTube + social) | Comms / Video | Gate | Currently done manually in the paper's pipeline, though the costing workbook rates this task as automatable (★ TECH) - worth reconciling; flagged, not yet resolved. |
| 53 | Publish initiative web page | Coordinator | Gate | Mechanical publish action; a coordinator still reviews the page before it goes live (see Fulfilling the social contract). |
| 54 | Create initiative page on website | Comms / Platform | Gate | Templated page generation from case data. |
| 55 | Send thank-you / continued-participation message | Comms | Gate | Templated message, low risk to automate. |

| # | Event | Human version | Tech version | Reason for choice |
|---|---|---|---|---|
| 56 | Decide panel enrolment | Research team / PI | Research team / PI | A research judgement about which cases warrant ongoing investment (see Further Enrichment with panelisation). |
| 57 | Retrieve filtered extract for analysis | Coordinator | Gate | Mechanical, rule-based retrieval against a whitelist. |
| 58 | Freeze snapshot for analysis | Scientific Manager | Gate | A mechanical freeze operation once the retrieval scope is set. |
| 59 | Control access; log data transfer | Coordinator | Gate | Access control against a whitelist is systematic and auditable. |

## Appendix 3. Structure and contents of a case folder

| Component | Contents |
|---|---|
| 1. Interview audio & video | Raw recordings of the visit (in-depth interview, filmed facility tour) - immutable originals |
| 2. Audio transcription | Transcript files: speaker-attributed, time-stamped, versioned |
| 3. Debrief | Debrief-session recording and its transcript, versioned |
| 4. Consent forms | Signed GDPR consent and image-rights authorisations |
| 5. Short video | Promotional clip and its social-media variants |
| 6. Supplementary material | Photos and documents provided by the initiative or scraped from its website |
| 7. Metadata & logs | Completion file; quality-and-completeness check; log of all interactions with the case; platform-page information; LATEST.json version pointer |
| 8. Analysis products | About-the-initiative summary, transaction grid, exploratory visuals and reports - versioned, each with a .json provenance sidecar |

## Appendix 4. Pipeline metrics extracted from the master Case Log

Source: "Procedure 3.1 – Appendix 1 – Case Log" (Master dashboard, 54 country sheets; snapshot exported 4 July 2026, last modified 25 June 2026. Extraction script: entry date = the case row's Year/Month columns; closure = "Year/Month of Closing the Case"; status filter = "OK - Closed". Dates are recorded at month granularity. 54 cases carried over from the 2023 proof of concept lack a closing month and are excluded from duration statistics.

### A4.1 Volume and funnel

The dashboard lists 3,247 dated lead/case rows across all country sheets. 3,247 entered the master log as leads; 45% of leads passed validation; 731 cases were closed at snapshot (50% of validated), with conversion up to 67% in mature territories (Colombia).

### A4.2 Cycle time from log entry to closure

Median 2 months; mean 2.97; 70% of cases close within three months of entering the log. Distribution (months → cases):

| **Months** | **0** | **1** | **2** | **3** | **4** | **5** | **6** | **7** | **8** | **9** | **10** | **11** | **12** | **13** | **15** |
|---|---|---|---|---|---|---|---|---|---|---|---|---|---|---|---|
| **Cases** | 33 | 149 | 180 | 113 | 70 | 42 | 36 | 18 | 17 | 9 | 2 | 4 | 2 | 1 | 1 |

### A4.3 Cycle time by country (n ≥ 10 closed cases)

| Country | Closed cases (dated) | Median months |
|---|---|---|
| **Colombia** | 161 | 3 |
| **France** | 79 | 3 |
| **Argentina** | 59 | 3 |
| **Kenya** | 50 | 2.0 |
| **Denmark** | 45 | 4 |
| **Peru** | 41 | 2 |
| **Spain** | 36 | 2.0 |
| **Costa Rica** | 34 | 3.0 |
| **Mexico** | 29 | 1 |
| **Ecuador** | 28 | 2.5 |
| **South Africa** | 21 | 1 |
| **Great Britain** | 16 | 2.0 |
| **India** | 14 | 1.5 |
| **Morocco** | 14 | 3.0 |
| **17 other countries** | 50 | N/A |

### A4.4 Monthly closures

| Month | Cases closed | Note |
|---|---|---|
| **2024-03** | 2 | |
| **2024-04** | 4 | |
| **2024-05** | 10 | |
| **2024-06** | 9 | |
| **2024-07** | 4 | |
| **2024-08** | 4 | |
| **2024-09** | 20 | |
| **2024-10** | 10 | |
| **2024-11** | 20 | |
| **2024-12** | 16 | |
| **2025-01** | 15 | |
| **2025-02** | 43 | |
| **2025-03** | 39 | |
| **2025-04** | 65 | |
| **2025-05** | 101 | |
| **2025-06** | 110 | peak |
| **2025-07** | 58 | |
| **2025-08** | 31 | |
| **2025-09** | 29 | |
| **2025-10** | 27 | |
| **2025-11** | 24 | |
| **2025-12** | 28 | |
| **2026-01** | 7 | |
| **2026-02** | 1 | |

Annual totals: 99 (2024), 570 (2025), 8 (Jan–Feb 2026).

### A4.5 Person-days estimate (order of magnitude)

Fieldwork: the budgeting instrument allocates four and a half person-days per completed case, all-inclusive. The 686 corpus cases measured here therefore represent ≈ 3,100 contracted field person-days (≈ 14 person-years at 220 working days/year). This is a lower bound: it excludes scouting effort sunk into the majority of the ~4,000 leads that never became completed cases.

Central coordination: the central data-collection team (data-collection manager, secretariat and panel manager, three fieldwork managers) ≈ 5 FTE over the 24 months of active collection (March 2024 – February 2026) ≈ 2,200 person-days. Video editing: at the proof-of-concept rate of 5–7 hours per case, ≈ 700 edited videos ≈ 500–700 person-days.

Order of magnitude for the data-collection operation as a whole: ≈ 5,800–6,000 person-days (≈ 26–27 person-years), excluding the technical/AI team, the PIs, and Paris

IAS administrative support. These estimates use designed loads (budgeted days, contract rates), not time sheets.

Finally, note this evaluation does not include quality control, tech work, support personnel and the time of the PIs and Co-PIs.

## Appendix 5. The project workforce, by category

Compiled 4 July 2026, headcounts updated 29 August 2026, from the project's operational instruments: the Interviewers Tracker (snapshot 29 June 2026), the team organigram (Figure 4), and the 2024/2025 annual reports. Individual names are withheld; the full roster is held internally.

| Category | Headcount | Contract / affiliation | Geography |
|---|---|---|---|
| **Principal investigators** | 2 | Academic appointments | France/UK; Austria |
| **Co-investigators** | 3 | Academic appointments | Austria, USA, UK/France |
| **Investigation support (PA, direction assistant)** | 2 | Institutional staff | Austria, France |
| **Data collection manager** | 1 | Core-team contract | France (global remit) |
| **CTO** | 1 | Core-team contract | France |
| **AI expert** | 1 | Core team | France |
| **Secretariat** | 1 | Institutional staff (Paris IAS) | France |
| **Technical / validation collaborators (incl. interns)** | 8 | Service contracts, collaborations | various |
| **Fieldwork managers** | 3 | Core-team contracts | regional portfolios (Europe/Africa/Asia/USA; Latin America; Latin America/Asia) |
| **Interviewers - phase 1-2 tracker (incl. regional coordinators)** | 62 | Service contracts, via local subcontractors or direct | ~30 countries |
| **Interviewers - earlier phases (additional)** | 19 | PoC and phase-1 service contracts | ~12 countries |
| **Video editors / video-asset contractors** | 6 | Service contracts | France, Colombia, Hong Kong |
| **Scientific Advisory Board** | 7 | Advisory; two meetings per year | international |
| **Friends of the Socioscope (network)** | 14 | Informal network | international |
| **Regional data-collection partner organisations** | 8 | Service contracts under local labour law | Colombia/Peru/Argentina; UK/Africa; Costa Rica; Mexico; Kenya/Uganda; USA; Paraguay; South Africa |
| **Host / research institutions** | 3 | Paris IAS (host, admin, ethics); CSH Vienna; LSE | France, Austria, UK |
| **Funder** | 1 | Grant (NOMIS Foundation) | Switzerland |

Reading note: the operational centre of gravity is small - a dozen core roles - while the field layer is wide (81 interviewers across two phases) and contractually mediated by eight regional partner organisations, consistent with the cost structure described in Variations and costs of the pipeline.

## Appendix 6. Project personnel

Credit to persons engaged on the project side across the proof-of-concept (2022–23) and phase 1–2 collection (2024–26): investigators, management, field, technical, editorial and advisory roles. Interviewees and case participants are not listed.

### Principal investigators and co-investigators

- Saadi Lahlou - Principal Investigator, Global
- Helga Nowotny - Principal Investigator, Global
- Stefan Thurner - Co-investigator
- Mirta Galesic - Co-investigator
- Paulius Yamin - Co-investigator
- Shriya Sekhsaria - Research & AI
- Johannes Oster - Research & AI

### Investigation support

- Barbara Grassauer - PA to H. Nowotny
- Irina Nosulenko - Direction assistant

### Central data collection team

- Juan Pablo Caicedo - Data Collection Manager, Global
- Valentine Fournand - Secretariat & Panel Manager

### Fieldwork managers

- Julia Movshovich - Fieldwork Manager, Europe, Africa, Asia, USA
- Sofía de Vega - Fieldwork Manager, Latin America, Spain
- María Mejía - Fieldwork Manager, Latin America, Asia

### Quality control

- Alex Cano - Quality controller (39-point completeness review)

### Technical team

- Antoine Cordelois - CTO
- Do Huynh - AI expert
- Yohann Gablowski - AI expert
- Eliot Boutherin - Developer
- Solène De Bonis - UX/UI design
- Annabelle Gouttebroze - Research intern; literature review study

- Julian Madera - Research intern; literature review study
- Kayode Adeniyi - Research intern, Developer
- Jinyu Cong - intern, NLP evaluation; LLM validation study (2024)
- Jeevya Aroun - intern, NLP evaluation; LLM validation study (2024)

**Proof-of-concept team (2022–23)**

- Loïc Bonin - Fieldwork manager assistant (to Emmanuelle Honoré), proof-of-concept phase; process documentation
- Emmanuelle Honoré - Fieldwork Manager and Interviewer, France (proof-of-concept phase); Interviewer, USA (2025)

**Contractors (video, field services)**

- Hugo Saugier - Video editor (PoC and after)
- Angélica María Pico Pedraza - Video assets (thumbnails, YouTube uploads), Colombia (Bucaramanga)
- Karen Garcia – Video editing
- Celine Gurteen – Video editing
- Juliette Capus – Video editing
- Ricardo Talens - Video editing (case videos)

**Interviewers and regional coordinators (phase 2 tracker)**

- Julian May - Regional Coordinator & interviewer, South Africa
- Romina Sarmiento - Regional / Field Coordinator + Interviewer, Argentina, Latam
- Astrid Frank-Bojsen - Meso-structure casework + interviews, Denmark, UK
- Mercedes Ejarque - Interviewer, Argentina, Latam
- Milena Duran - Interviewer, Argentina, Latam
- Marina Baima - Interviewer, Argentina, Latam
- Lucia Sosa - Interviewer, Argentina, Latam
- Ignacio Bordoli - Interviewer, Argentina, Latam
- Ana María Mendoza - Interviewer, Colombia, Latam
- Daniela Pedroza - Interviewer, Colombia, Latam
- Nathaly Jiménez - Interviewer, Colombia, Latam
- Stephany Navas - Interviewer, Colombia, Latam
- Santiago Borda - Interviewer, Colombia, Latam
- Paola Abril Campos Rivera - Interviewer, Mexico, Latam
- Beatriz Michelle Ramírez Pérez - Interviewer, Mexico, Latam
- Alejandra González Moreno - Interviewer, Mexico, Latam
- Alejandra Maldonado Esquer - Interviewer, Mexico, Latam
- Finn Richardson - Interviewer, Costa Rica, Latam

- Pablo Hellmund - Interviewer, Costa Rica, Latam
- Bérénice Perroud - Interviewer, France, Switzerland, Denmark
- Saadi Lahlou - Interviewer, France, Colombia, Denmark
- Helga Nowotny - Interviewer, France
- Juan Pablo Caicedo - Interviewer, Colombia, Peru, Lithuania
- Valentine Fournand - Interviewer, USA
- Paulius Yamin - Interviewer, Colombia, Lithuania
- Oleg Dremov - Interviewer, Georgia, Armenia, Israel
- Ganesh Radha-Udayakumar - Interviewer, India
- Cathy Kamanu - Regional / Field Coordinator + Interviewer, Kenya
- Brahim Lebbar - Interviewer, Morocco
- Lamia Afrasse - Interviewer, Morocco
- Lucie Friedrich - Interviewer, France, Europe
- Matej Vohryzek - Interviewer, New Zealand, Tahiti, Japan
- Samara Zuckerbrod - Interviewer, Wales
- Boris Pun - Interviewer, Hong Kong
- Anthony Chun Fung Cheung - Interviewer, Hong Kong
- Leonardo Herrera - Regional / Field Coordinator + Interviewer, Peru, Latam
- Alvaro Elorrieta - Interviewer, Peru, Latam
- Andrea Rivera - Interviewer, Peru, Latam
- Eizo Muñoz - Interviewer, Peru, Latam
- Judith Huamani - Interviewer, Peru, Latam
- Fabian Rojas - Interviewer, Peru, Latam
- Karla Diaz - Interviewer, Peru, Latam
- Gianina Chávarry Minaya - Interviewer, Peru, Latam
- Gabriela Salazar - Interviewer, Peru, Latam
- Lucía Rucoba - Interviewer, Peru, Latam
- Malgorzata Patok - Interviewer, Poland, France
- Annya Crane - Interviewer, Spain, Netherlands
- Andrea Padilla - Interviewer, UK, Nigeria, Senegal, Sri Lanka
- Eoghan McDonaugh - Interviewer, UK, Africa
- Emmanuelle Hopkins - Interviewer, USA
- Ravenn Tripplet - Regional / Field Coordinator + Interviewer, USA
- Betania Ayala Valdez - Interviewer, Paraguay, Latam
- Gustavo Grajeda - Interviewer, Paraguay, Latam
- Keshia Lauren Anne Hoaeane - Regional / Field Coordinator + Interviewer, South Africa, Africa
- Lance Scheepers - Interviewer, South Africa, Africa
- Taiye Orija - Interviewer, South Africa, Africa

- Siliziwe Zote - Interviewer, South Africa, Africa
- Asemahle Mali - Interviewer, South Africa, Africa
- Jen Horn - Interviewer, Philippines, Asia
- Joshua Kurniawan - Interviewer, Indonesia, Asia
- L.A. Bretous - Interviewer, USA
- Jeremy Barajas - Interviewer, USA
- Mariela Dyer - Interviewer, USA
- Fred Munene - Interviewer, Kenya
- Sharon Amali - Interviewer, Uganda
- Denis Woniala - Interviewer, Uganda
- Anne Koigi - Interviewer, Kenya

**Interviewers (proof-of-concept phase)**

- Adéla Vašků - Interviewer, Czechia
- Albena Shkodrova - Interviewer, Bulgaria
- Alvaro Elorrieta - Interviewer, Peru
- Amélie Peuteuil - Interviewer, France
- Asemahle Mali - Interviewer, South Africa
- Camille Senepin - Interviewer, France
- Camilo Ordóñez - Interviewer, Colombia
- Christine Adongo - Interviewer, Kenya
- Clémentine Decroix - Interviewer, France
- Davide Gnoato - Interviewer, Italy
- Denis Woniala - Interviewer, Uganda
- Fred Munene - Interviewer, Kenya
- Hélène Peters Zwingelstein - Interviewer, France
- Ho Jack Yong - Interviewer, Singapore
- Jérémie Szlamowicz - Interviewer, France
- Lauriane Dos Santos - Interviewer, Cook Islands, French Polynesia
- Loic Bonin - Interviewer, Denmark
- Mei Anne Hills - Interviewer, UK
- Nicole Alexander - Interviewer, Denmark, UK; co-author of the Danish-cases study (with A. Frank-Bojsen)
- Sarah Baudry - Interviewer, France
- Sifa Florence Sangwa - Interviewer, Rwanda
- Tony Alfred - Interviewer, Tanzania
- Umberto Cao - Interviewer, France

**Scientific Advisory Board (membership during Phase 1)**

- Catherine Bassani - SAB member
- Arnold Tukker - SAB member
- Reinhilde Veugelers - SAB member
- Joachim von Braun - SAB member
- Patrick Caron - SAB member
- Deshen Moodley - SAB member
- Olivier Bouin - SAB member

**Friends of the Socioscope**

- Erik Arnold - Network member
- Linda Bell - Network member
- Nicolas Bricas - Network member
- Wolfgang Burtscher - Network member
- Mathias Dewatripont - Network member
- Jakob Edler - Network member
- Bruno Herault - Network member
- François Jegou - Network member
- David Kanter - Network member
- Bettina Laville - Network member
- Daniel Nairaud - Network member
- Sophie Nicklaus - Network member
- David Siaussat - Network member
- David Stark - Network member

## Appendix 7. Contents of the Data Collection Protocol

The protocol is the single instrument that makes cases comparable across countries, and it is the piece other teams most often ask to see. It is not reproduced here: it runs to fifty-five pages and is available from the authors on reasonable request.

Its table of contents is given below (V8, January 2025, 55 pages), to show what a field-ready LSQR protocol has to cover.

The protocol is used with four training videos, produced in 2024 and watched by every interviewer before the first case: Training 1, project description, logs and contact phase (8 min 47 s); Training 2, preparing your interview (8 min 37 s); Training 3, executing the interview (26 min 55 s); Training 4, finalising your mission (6 min 58 s). Together they run to 51 minutes and cover the same three stages as the protocol, showing the procedures being performed rather than described.